\documentclass[useAMS,usenatbib]{mn2e}
\usepackage{fixltx2e}
\usepackage{amsmath}
\usepackage{graphicx}
\usepackage{subfig}
\usepackage{float}
\usepackage{upquote}
\usepackage{footnote}

\title[Dynamics of stars in a barred-spiral galaxy]{Dynamics of stars around spiral arms in an N-body/SPH simulated barred-spiral galaxy}
\author[Grand et al.]
{Robert J.J. Grand $^1$\thanks{rjg2@mssl.ucl.ac.uk}, Daisuke Kawata $^1$, Mark Cropper $^1$\\
$^1$ Mullard Space Science Laboratory, University College London, Holmbury St. Mary, Dorking, Surrey, RH5 6NT}

\pagerange{\pageref{firstpage}--\pageref{lastpage}} \pubyear{2011}

\begin{document}

\label{firstpage}
\maketitle

\begin{abstract}
We run N-body smoothed particle hydrodynamics (SPH) simulations of a Milky Way sized galaxy. The code takes into account hydrodynamics, self-gravity, star formation, supernova and stellar wind feedback, radiative cooling and metal enrichment. The simulated galaxy is a barred-spiral galaxy consisting of a stellar and gas disc, enveloped in a static dark matter halo. Similar to what is found in our pure N-body simulation of a non-barred galaxy in \citet{GKC11}, we find that the spiral arms are transient features whose pattern speeds decrease with radius, in such a way that the pattern speed is similar to the rotation of star particles. Compared to the non-barred case, we find that the spiral arm pattern speed is slightly faster than the rotation speed of star particles: the bar appears to boost the pattern speed ahead of the rotational velocity. We trace particle motion around the spiral arms at different radii, and demonstrate that there are star particles that are drawn towards and join the arm from behind (in front of) the arm and migrate toward the outer (inner) regions of the disc until the arm disappears as a result of their transient nature. We see this migration over the entire radial range analysed, which is a consequence of the spiral arm rotating at similar speeds to star particles at all radii, which is inconsistent with the prediction of classical density wave theory. The bar does not prevent this systematic radial migration, which is shown to largely preserve circular orbits. We also demonstrate that there is no significant offset of different star forming tracers across the spiral arm, which is also inconsistent with the prediction of classical density wave theory. 
\end{abstract}


\section{Introduction}

The most well known theory of spiral arm structure is the so-called spiral density wave theory, which describes the spiral arms as quasi-stationary density waves (\citealt{L60}; \citealt{LS64}) that rotate with the same pattern speed at every radius. The spiral arm can therefore be described by a wave function. To describe the spiral arm with a wave function is beneficial because it is then possible to extract analytic solutions of quantities such as the dispersion relation by use of Euler's equations of motion. For these solutions, the tight winding approximation is made, which require very tight spirals whose pitch angles remain constant (see \citealt{A84} and references therein). Spiral galaxies have been discussed in the context of global density wave patterns (e.g. \citealt{AL97}; \citealt{YCG02}; \citealt{AF11}; \citealt{LR11}). While some studies \citep{DT94} argue the existence of long-lived patterns in their simulations, the long-lived classic spiral density structure has never been reproduced self-consistently, and in all numerical simulations, the spiral arm is seen to be a transient structure \citep{Se11}.  

The transient property exhibited by simulated spiral arms has stimulated the emergence of new discussion of spiral arm evolution. A recent suggestion includes multiple wave modes of different pattern speeds that create the transient nature of spiral arms by constructively and destructively interfering with one another, thereby ensuring the growth and decay of the stellar density enhancement (e.g. \citealt{SK91} \citealt{MT97}; \citealt{MQ06}; \citealt{RD11}; \citealt{QDBM10}). 

Another possible explanation of the transient nature is the co-rotating spiral arm. Here, the spiral arm is considered to be rotating with the material at every radius. Naturally, the transient property of the arm is manifested by the winding, which propagates to breaks and bifurcations of the spiral arm. Such breaks in the spiral arm structure were found in \citet{WBS11} and \citet{GKC11}, who performed high resolution N-body smoothed particle hydrodynamics (SPH) simulations (and in the latter case, pure N-body simulations) of an isolated spiral galaxy. They show that the pattern speed of the spiral arms decreases with radius, such that it follows the circular velocity of star particles. 

The co-rotating spiral arm is found to have significant consequences on the radial migration of star particles. \citet{GKC11} demonstrated a new type of systematic motion of star particles close to the spiral arm in their simulation, that leads to large, efficient radial migration of star particles all along the arm as opposed to the currently considered case of a single co-rotation radius (\citealt{LBK72}; \citealt{SB02}). The star particles were shown to join the arm from both sides. Star particles behind (in front of) the arm were accelerated (decelerated) continually because the similar rotation speeds of the star particles and spiral arm allowed the migrating star particles to stay very close to the density enhancement of the spiral arm. This mechanism is responsible for the steady gain/loss of angular momentum of the migrating star particles, whereby the star particle is allowed to find a new equilibrium in a higher/lower energy circular orbit, without scattering kinematically. The star particles never crossed the arm as they migrated, and stopped migrating when the high amplitude of the density enhancement disappeared owing to the transient nature of the arms. Because this simulation was an N-body simulation of a pure stellar disc with no bar or bulge, the only factor that could be responsible for the observed motion was the spiral arm features. 

Complementary to numerical simulations, observational tests of the pattern speed have been made by \citet{SW11}, who use the solutions to the Tremaine-Weinberg equations \citep{TW84} to perform a statistical analysis on NGC 1365, and find the best solution to be a pattern speed that decreases as $1/r$ (see also \citealt{MRM05}; \citealt{MRM06}; \citealt{MRM08}; \citealt{MRM09}). Another test is the presence of (or lack of) a clear offset between different star forming tracers across a spiral arm. In the context of density waves, star particles should flow through the spiral arm (everywhere except at a co-rotation radius) if the pattern speed is constant. As gas flows into the spiral arm from behind the arm inside co-rotation and from in front of the arm outside co-rotation, it is compressed into molecular clouds. Stars are born from molecular clouds and age as they continue to flow relative to the arm, leaving a clear trail of stellar evolution (star forming tracers) from one side of the arm to the other. This has recently been tested for NGC 4321 by \citet{FCK12}, who find no apparent offset between H$_{\alpha}$ and UV sources. \citet{FR11} also find no offset in twelve nearby spiral galaxies by observing the star forming tracers HI and CO, 24$\mu$m emission and UV emission to trace atomic gas, molecular gas, enshrouded stars and young stars respectively. Both studies are evidence against long-lived spiral arms. 

In this study, our aim is to build upon our previous study \citep{GKC11} that focused on N-body dynamics in a pure stellar disc, and extend this research on the spiral arm pattern speed and star particle dynamics in high resolution N-body/SPH simulations of a barred spiral galaxy. This will enable us to study the spiral arm and its effects in a more realistic context, and to determine whether or not the presence of gas, star formation and a bar produces any significant effect on particle motion that may be distinguished from those seen in the pure N-body simulation. In comparison to \citet{GKC11}, we present a more robust method for determining the apparent pattern speed of the spiral arm, and attention is given to the energy evolution of particles that undergo radial migration at many radii. Additionally, we show the distribution of young star particles of different ages to check for offsets in different star forming tracers. Although the analysis could be extended to the bar region, this paper focuses on the spiral arm. Hence we leave the analysis of the structure and evolution of bars to future studies.

In Section 2, a description of the SPH code is given before the model set up and the chosen initial parameters are outlined. In Section 3 we present the results of our analysis, compare them with previous studies and discuss their implications. In Section 4 we summarise the significance of the results and remark upon the value of the simulations and future work.

\section{Method and Model Setup}

\subsection{GCD+ code}

In our simulation, we use an updated version of the original galactic chemodynamical evolution code, GCD+, developed by \citet{KG03}. A detailed description of the code is seen in \citet{RK11}. Here we give a brief outline. GCD+ is a three-dimensional tree N-body/SPH code (\citealt{GM77}; \citealt{L77}; \citealt{BH86}; \citealt{HK89}; \citealt{KWH96}) that incorporates self-gravity, hydrodynamics, radiative cooling, star formation, supernova feedback and metal enrichment. This latest version of GCD+ takes into account metal diffusion as suggested by \citet{GGB09}. The scheme follows that of \citet{RP07}: we use their artificial viscosity switch (Morris 1997) and artificial thermal conductivity to resolve Kelvin-Helmholtz instabilities \citep{KO09}. Further adaptations include those of adaptive softening \citep{PM07} and an individual time step limiter \citep{SM08} in order to correctly resolve particle response to shock layers ploughing through material from supernova and wind blown bubbles (e.g. \citealt{MBG10}; \citealt{DDV11}).  

Radiative cooling and heating is calculated with CLOUDY (v08.00: \citealt{FKV98}). UV background radiation is also taken into account \citep{HM96}. Our star formation formula corresponds to the Schmidt law. We set a threshold density for star formation, $n_{th}$, which means that star formation will occur for any region that exceeds this density and the velocity field is convergent.

We assume that stars are distributed according to the \citet{S55} initial mass function (IMF). Chemical enrichment by both Type II \citep{WW95} and Type Ia supernovae (\citealt{I99}; \citealt{KTN00}) and mass loss from intermediate-mass stars \citep{vdHG97} are taken into account. The new version of GCD+ uses a different scheme for star formation and feedback (see \citealt{RK11}). We now keep the mass of the baryon (gas and star) particles completely the same, unlike our old version \citep{KG03} or the majority of SPH simulations which include star formation. 

The main parameters that govern star formation and supernova feedback \citep{RK11} are set as follows: the star formation density threshold, $n_{th} = 1.0$  $\rm cm^{-3}$ ; star formation efficiency, $C_* = 0.1$ ; supernova energy input, $E_{\rm SN} = 10^{50}$ $\rm erg$ per supernova; and stellar wind energy input, $E_{\rm SW} = 10^{36}$ $\rm erg$ $\rm s^{-1}$. Each particle in the simulation is assigned a unique ID number. This makes it easy to trace any particle during the evolution of the simulation.  

\subsection{Simulation Setup}

Our simulated galaxy consists of a spherical static dark matter halo and two live discs: a stellar disc and a gas disc. The dark matter halo density profile follows that of \citet{NFW97}:

\begin{equation}
\rho _{\rm dm} = \frac{3 H_0^2}{8 \pi G} (1+z_0)^3 \frac{\Omega _0}{\Omega (z)} \frac{\rho _{\rm c}}{ cx(1+cx)^2} ,
\label{eq1}
\end{equation} 
where $\rho _{\rm c}$ is the characteristic density described by \citet{NFW97}, the concentration parameter, $c = r_{\rm 200}/r_{\rm s}$, and  $x= r/r_{\rm 200}$. The scale length is $r_{\rm s}$, and $r_{\rm 200}$ is the radius inside which the mean density of the dark matter sphere is equal to 200$\rho _{\rm crit}$ (where $\rho _{crit} = 3 H_0^2 / 8 \pi G$; the critical density for closure):

\begin{equation}
r _{200} = 1.63 \times 10^{-2} \left( \frac{M_{200}}{h^{-1} M_{\odot}} \right)^{\frac{1}{3}} \left[\frac{\Omega _0}{\Omega (z_0)}\right]^{-\frac{1}{3}} (1+z_0)^{-1} h^{-1} \rm kpc.
\end{equation} 
We assume $M_{200} = 1.5 \times 10^{12}$ $\rm M_{\odot}$, $c=10$, $\Omega _0 = 0.266$, $z_0=0$ and $H_0=71$ $\rm km$ $\rm s^{-1}$ $\rm Mpc^{-1}$. 

\begin{figure}
\centering
\includegraphics[scale=0.43]{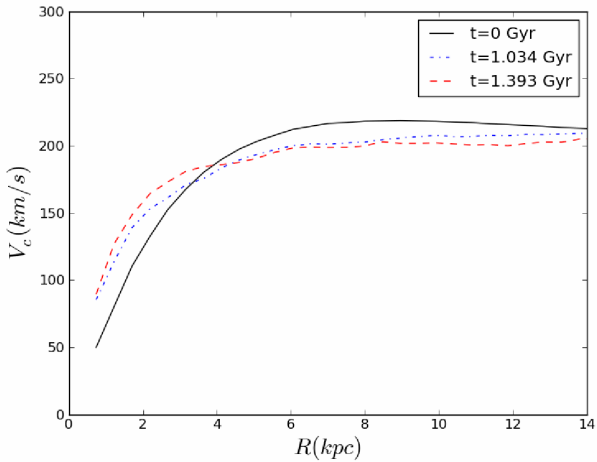}
\caption[rotational velocity at t$=1.77$ Gyr, and initial rotation curve.]
{The rotational velocity at $t=0$ (solid black line), at t$=1.034$ (dot-dashed blue line) and $t=1.393$ Gyr (dashed red line).}
\label{vcrotc}
\end{figure}

\begin{figure*}
\begin{center}

  \subfloat{\includegraphics[scale=0.41]{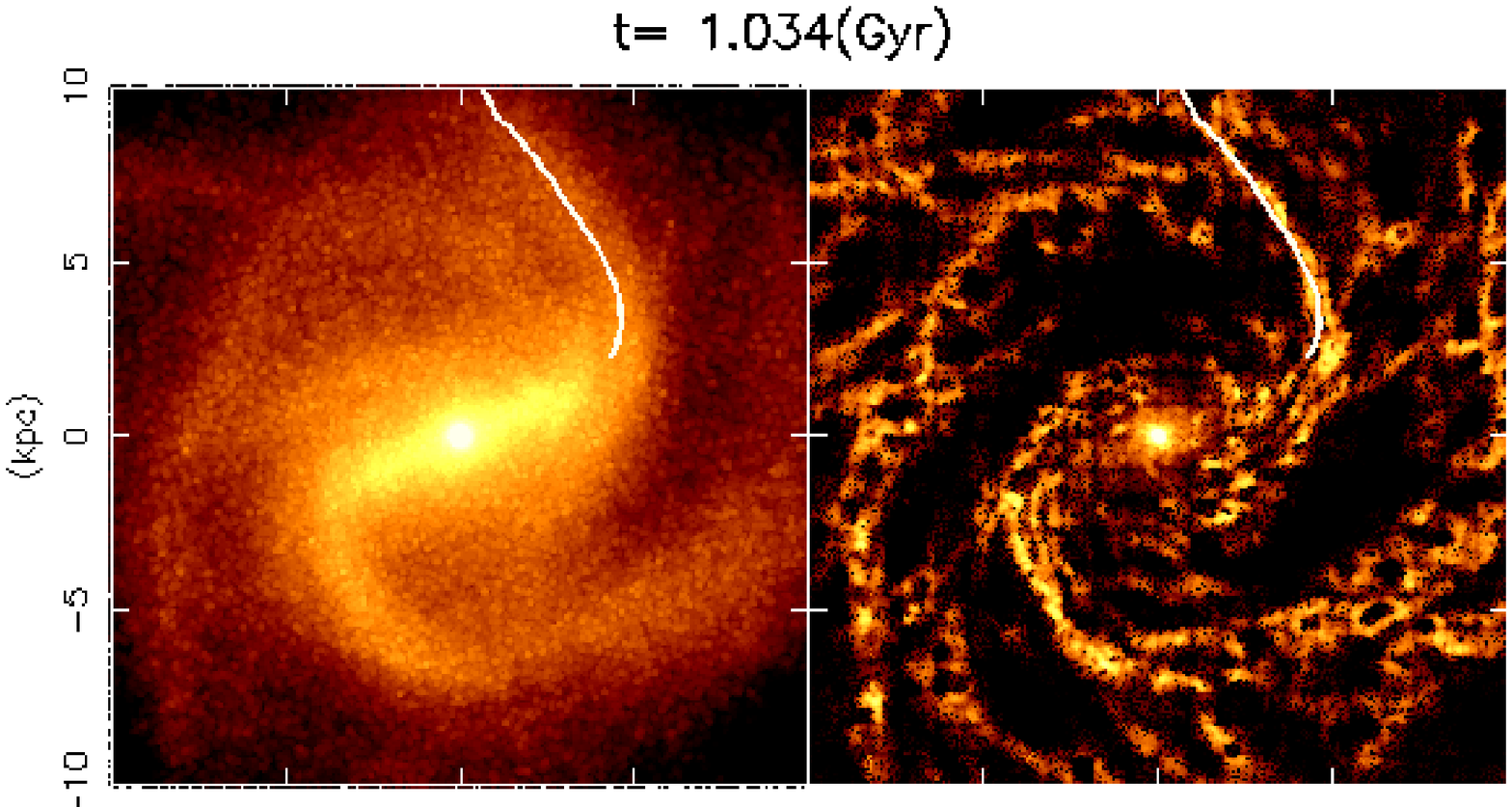}} \vspace{-0.0mm}
  \subfloat{\includegraphics[scale=0.41] {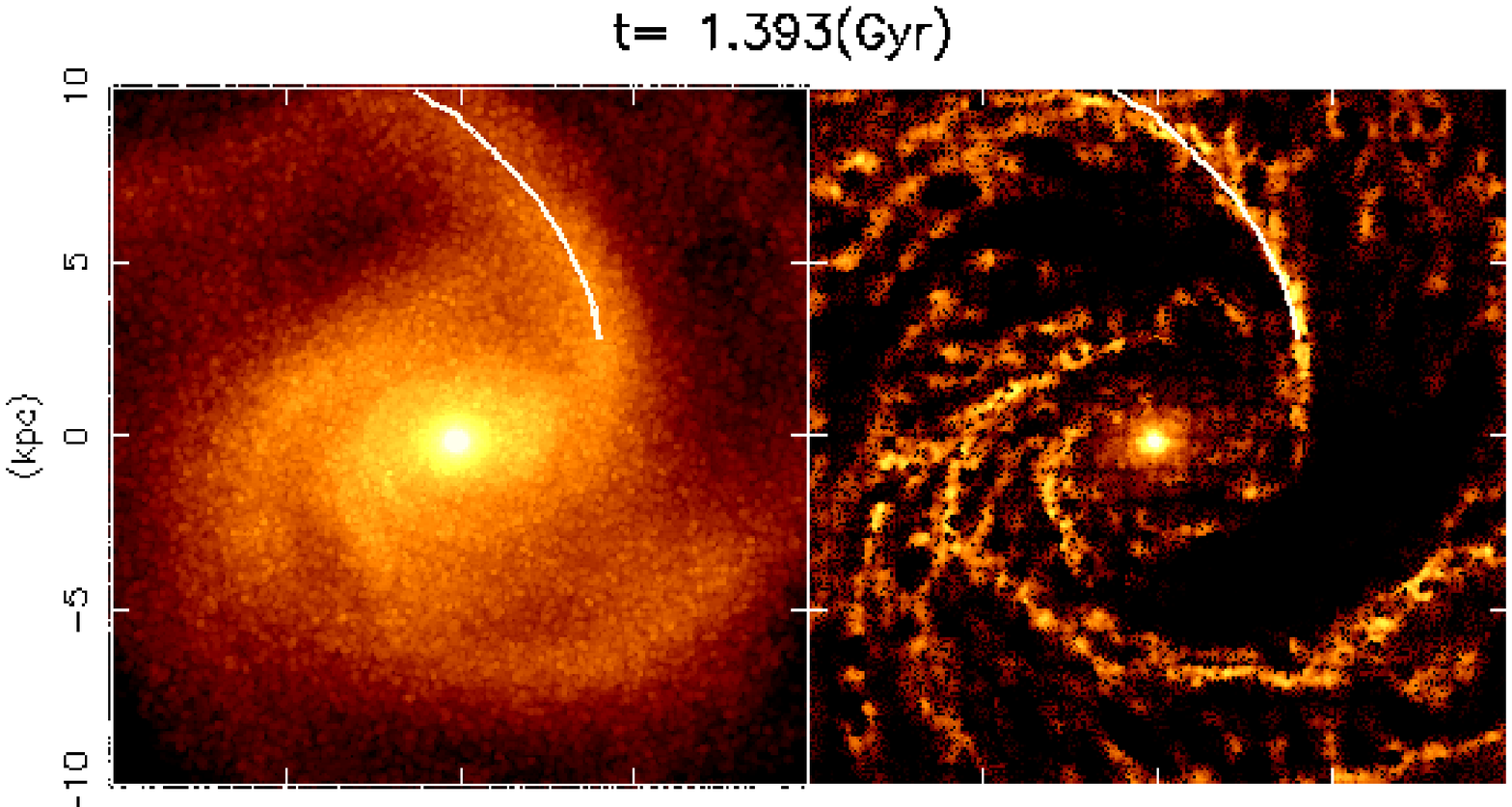}} \\

\caption[]{Snapshots of the face-on view of the simulated galaxy at $t = 1.034$ Gyr (left) and $t = 1.393$ Gyr (right). The left images show the stellar density map, and the right images show the gas density map. The bar is strong at the earlier time, and becomes smaller at the later time.}
\label{galprev}
\end{center}
\end{figure*} 

\begin{figure}
\centering
\includegraphics[scale=0.43]{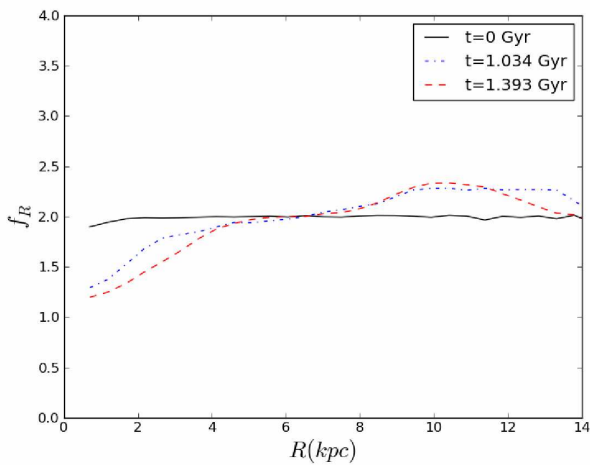}
\caption[The ratio of velocity dispersions in the radial and $z$ direction, at $t$ = 0 (solid black line), t$=1.034$ (dot-dashed blue line) and $t=1.393$ Gyr (dashed red line) plotted as a function of radius.]
{The ratio of velocity dispersions in the radial and $z$ direction, at $t$ = 0 (solid black line), t$=1.034$ (dot-dashed blue line) and $t=1.393$ Gyr (dashed red line) plotted as a function of radius.}
\label{frQ}
\end{figure}

\begin{figure}
\centering
\includegraphics[scale=0.43]{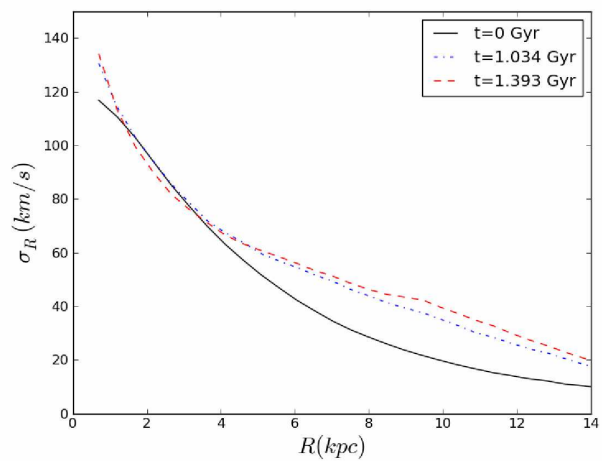}
\caption[]
{Radial velocity dispersion computed at $t$ = 0 (solid black line) t$=1.034$ (dot-dashed blue line) and $t=1.393$ Gyr (dashed red line), as a function of radius.}
\label{sigr}
\end{figure}

\begin{figure}
\centering
\includegraphics[scale=0.43]{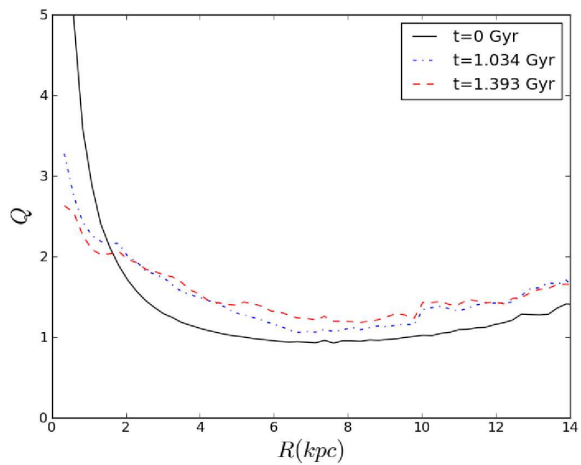}
\caption[]
{Toomre's instability parameter, $Q$, computed at $t$ = 0 (solid black line) t$=1.034$ (dot-dashed blue line) and $t=1.393$ Gyr (dashed red line), as a function of radius.}
\label{newQ2}
\end{figure}

The stellar disc is assumed to follow an exponential surface density profile:

\begin{equation}
\rho _{\rm d,*} = \frac{M_{\rm d,*}}{4 \pi z_{\rm d,*} R_{\rm d,*}} {\rm sech}^2 \left(\frac{z}{z_{\rm d,*}}\right) {\rm exp}\left(-\frac{R}{R_{\rm d,*}}\right),
\end{equation} 
where the disc mass, $M_{\rm d,*} = 5 \times 10^{10}$ $\rm M_{\odot}$, the scale length, $R_{\rm d,*} = 2.5$ kpc and the scale height $z_{\rm d,*} = 0.35$ kpc, which is constant over the disc. The velocity dispersion for each three dimensional position of the disc is computed following \citet{SMH05} to construct the almost equilibrium initial condition. One free parameter in this method is the ratio of the radial velocity dispersion to the vertical velocity dispersion, $f_R$, which relates as $f_R = \sigma _R / \sigma _z$. We choose $f_R = 2$ in the simulation shown. The initial circular velocity curve is shown in Fig. \ref{vcrotc} (solid line). The initial rotation speed at 8 kpc is $\sim 220$ $\rm km$ $\rm s^{-1}$. 

The gas disc is set up following the method described in \citet{SMH05}. The radial surface density profile is assumed to follow an exponential law like the stellar disc. The initial vertical distribution of the gas is iteratively calculated to be in hydrostatic equilibrium assuming the equation of state calculated from our assumed cooling and heating function. For the gas disc, we set the disc mass, $M_{\rm d,g} = 1 \times 10^{10}$ $\rm M_{\odot}$, the scale length, $R_{\rm d,g} = 4.0$ kpc.

\begin{figure}
\centering
\includegraphics[scale=0.43]{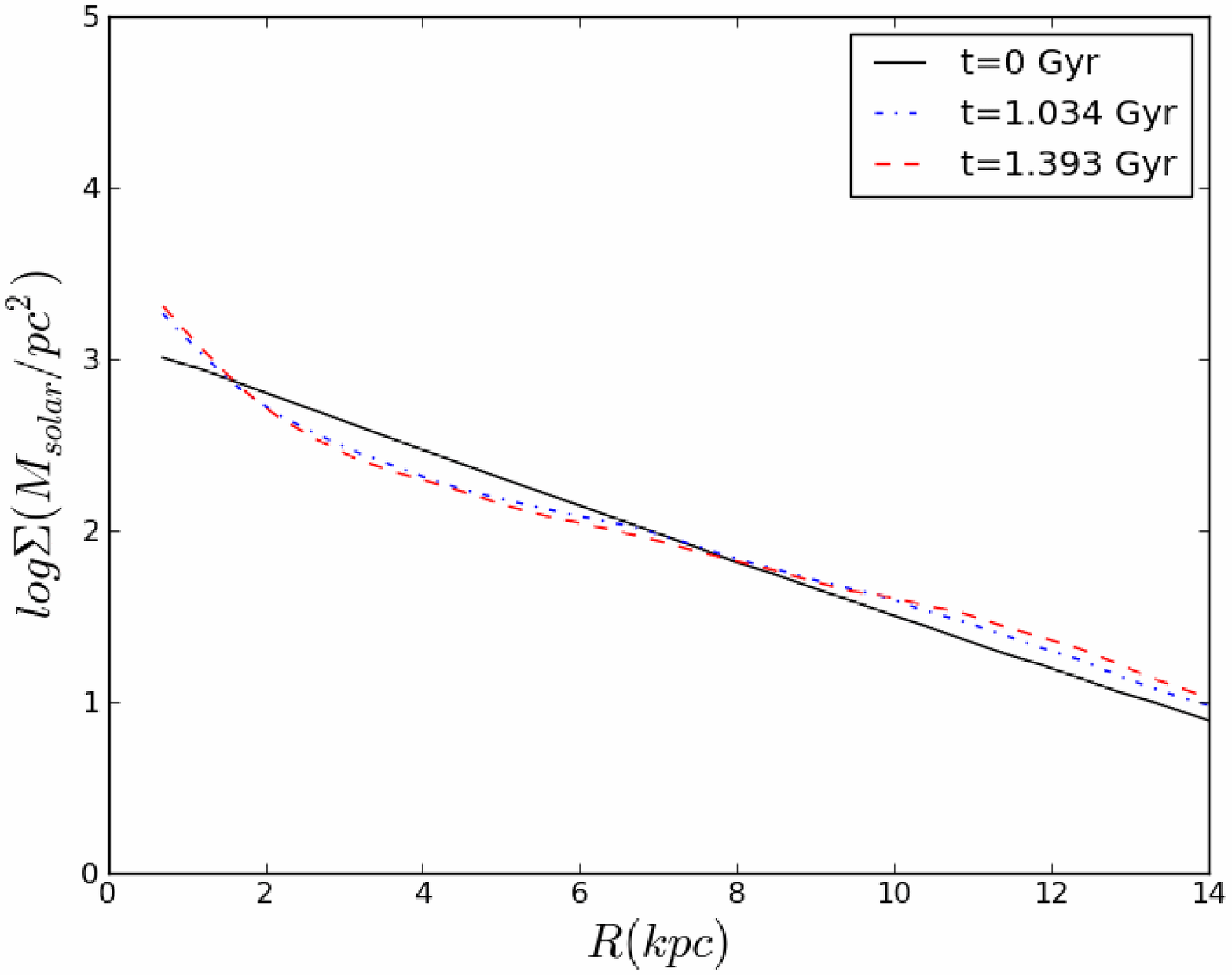}
\caption[]
{Logarithm of the surface density, computed at $t$ = 0 (solid black line) and t$=1.034$ (dot-dashed blue line) and $t=1.393$ Gyr (dashed red line), as a function of radius.}
\label{Sig}
\end{figure}

We use $N_{\rm d,*} = 2 \times 10^6$ and $N_{\rm d,g} = 4 \times 10^5$ particles for stars and gas respectively, and therefore the mass of each particle is $2.5 \times 10^4$ $\rm M_{\odot}$. \citet{Fu11} show that if more than one million particles are used to describe the disc component, artificial heating that suppresses the spiral arm formation is not significant. Our simulation uses a total of $2.4 \times 10^{6}$ particles and therefore is expected to be less affected by artificial heating. We adopt a softening length equal to the smoothing length but set the minimum softening length to 340 pc for gas particles and apply a fixed softening length of 340 pc for star particles, with the spline softening suggested by \citet{PM07}. These  parameters of the stellar component are similar to that of the non-barred spiral galaxy simulated in \citet{GKC11}, but with a higher disc to halo mass ratio. To induce spontaneous bar formation (e.g. \citealt{OP73}), we have applied a lower concentration parameter, $c = 10$, in eq. (1).

\section{Results and Discussion}

The simulation set up in Section 2 was evolved for about 2 Gyr. The stellar and gas component is shown at two different times in Fig. \ref{galprev}, and we see a prominent bar spiral structure in both components. The strong bar develops around $t = 1.034$, and settles to a smaller bar before $t = 1.393$ Gyr. Similar to previous studies described in Section 1, we also find that the disc develops transient and recurrent spiral arms. In this paper, we focus our analysis on spiral arms at an early and late epoch in the evolution of the simulated galaxy, highlighted in Fig. \ref{galprev}. These times are referred to as the centre of each epoch throughout the paper. Particular attention is paid to these spiral arms because they are prominent arms, which facilitate our analysis and we are able to extract and more clearly demonstrate the key features that we want to identify, namely the pattern speed and the particle motion around the spiral arm.

The circular velocity at $t = 0$ and $t=1.034$ Gyr (early epoch) and $t=1.393$ Gyr (late epoch) is shown in Fig. \ref{vcrotc}. The circular velocity in the inner region after $t=0$ is significantly different from the initial circular velocity, owing to the strong gravitational field created by the developed bulge. Fig. \ref{frQ} shows $f_R$ as a function of radius at the same time steps. The value drops with time in the inner radii (bar region). Outside $R \sim 5$ kpc, $f_R$ increases slightly as the disc is heated by strong spiral structure, which increases the velocity dispersion, $\sigma _R$, shown explicitly in Fig. \ref{sigr}. The effect on spiral structure is quantified in Fig. \ref{newQ2}, which shows an increase of Toomre's instability parameter, $Q = \sigma _R \kappa / 3.36 G \Sigma _*$, in the spiral region, where $\kappa$ is the epicycle frequency and $\Sigma _*$ is the surface density of the stellar component. This is contrary to the bar region where $Q$ is lowered owing to the large surface density excess in the central region shown in Fig. \ref{Sig}. A bulge that creates this excess of central density are likely formed through secular evolution caused by the bar (e.g. \citealt{PN90}; \citealt{KK04} and references therein). The developed bulge is apparent in Fig. \ref{galprev}. 

We present analysis and discussion of two spiral arms at an early ($t \sim 1.034$ Gyr) and a later ($t \sim 1.393$ Gyr) epoch of the galaxy\textquotesingle s evolution. This is because the bar is strong at the early epoch, in contrast with the later epoch when the bar is comparatively weak. To quantify the bar strength, we use a gravitational force field method (e.g. \citealt{BB01}; \citealt{BVs05}). We first define a circular grid that covers an azimuth range of $0$ to $2 \pi$ and a radial range of $1$ kpc to $5$ kpc. At the centre of each grid point, the radial and tangential forces are calculated, which are then used to calculate the ratio:

\begin{equation}
Q_T (R,\theta) = \frac{|F_T(R,\theta)|}{\bar{F}_R(R,\theta)} ,
\end{equation} 
where $F_T(R, \theta)$ is the tangential force at a given grid point of coordinates $(R, \theta)$, and $\bar{F}_R(R, \theta)$ is the mean radial force averaged over each azimuth at a given radius \citep{CS81}. A maximum, $Q_{b,i}$, is found in each quadrant, where quadrants $i = 1, 2, 3$ and $4$ are defined by setting the major and minor axes of the bar to the $x-$ and $y-$axes respectively. The bar strength is then defined as the average of these four values: $Q_b = \sum_{i=1}^{4} Q_{b,i} / 4$. At the early epoch, $Q_b = 0.27$, and at the late epoch, $Q_b = 0.11$. According to the classification scheme outlined in \citet{BB01}, these values correspond to a class 3 and class 1 bar at the early and late epochs respectively.

First we present the analysis and results of the pattern speeds of the chosen spiral arms. Then we examine the motion of selected particles around the arm, and present and discuss an analysis of their angular momentum and energy evolution. We compute the angular momentum evolution around both spiral arms, and make a comparison between each case. We also examine the position of star particles of different ages in and around the spiral arm, which would be an observational test for pattern speeds of spiral arms (\citealt{Ta08}; \citealt{EG09}; \citealt{FR11}; \citealt{FCK12}). If the spiral arms rotate with a constant pattern speed, systematic offsets in azimuth between age populations and the spiral arm as a function of radii are expected.

It should be noted however, that we also applied similar analyses to other spiral arms that developed at different times in this simulation as well as spiral arms in other barred spiral simulations with different initial configurations of the disc and dark matter halo. We find that all the spiral arms we analysed show very similar results to those shown in this section (see also \citealt{KGC11}).

\begin{figure*}
\begin{center}

  \subfloat{\includegraphics[scale=0.65] {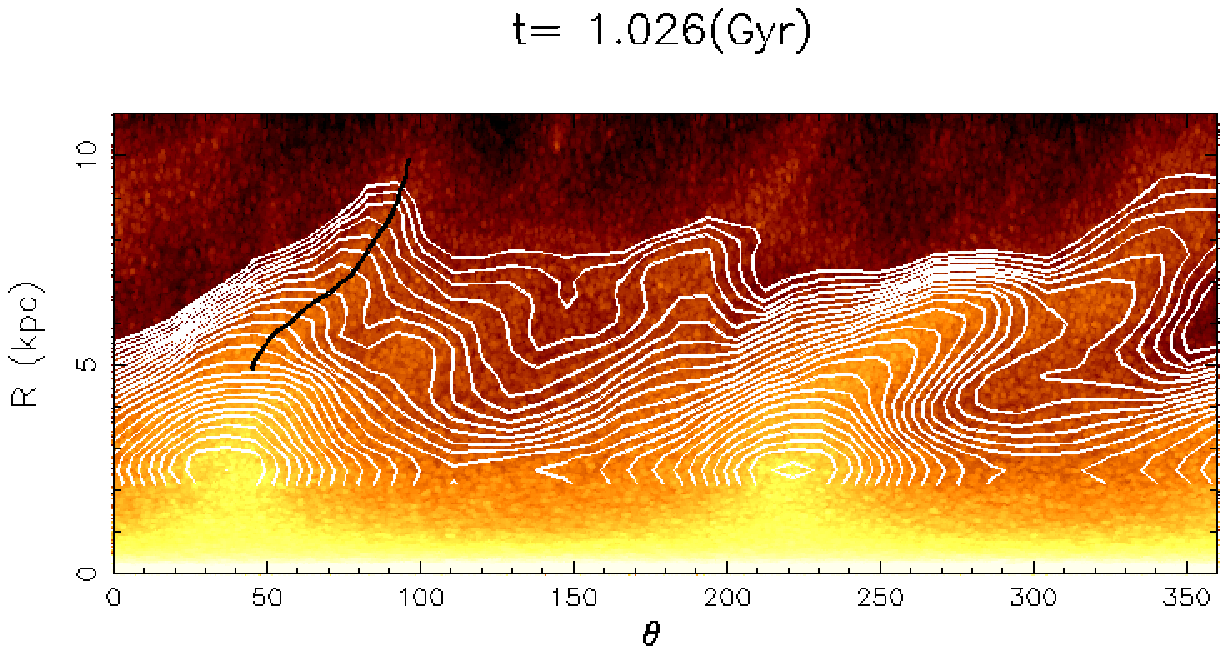}} \hspace{-11.0mm}
  \subfloat{\includegraphics[scale=0.65] {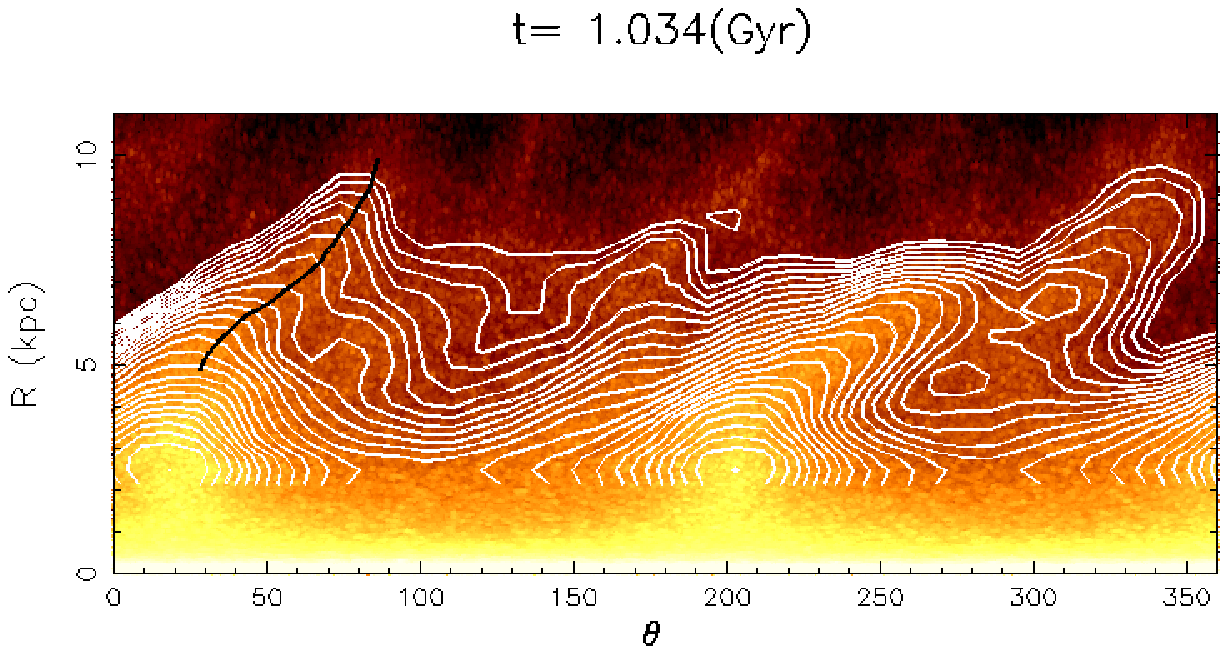}} \\ \vspace{-10.5mm}
  \subfloat{\includegraphics[scale=0.65] {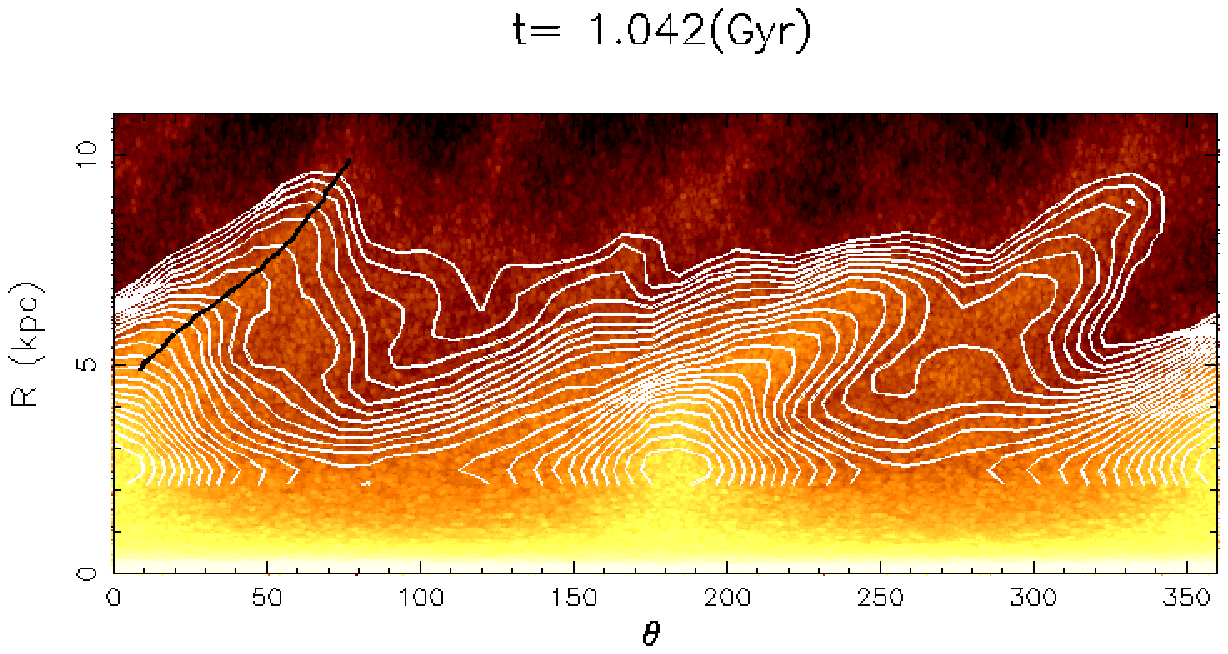}} \hspace{-11.0mm}
  \subfloat{\includegraphics[scale=0.65] {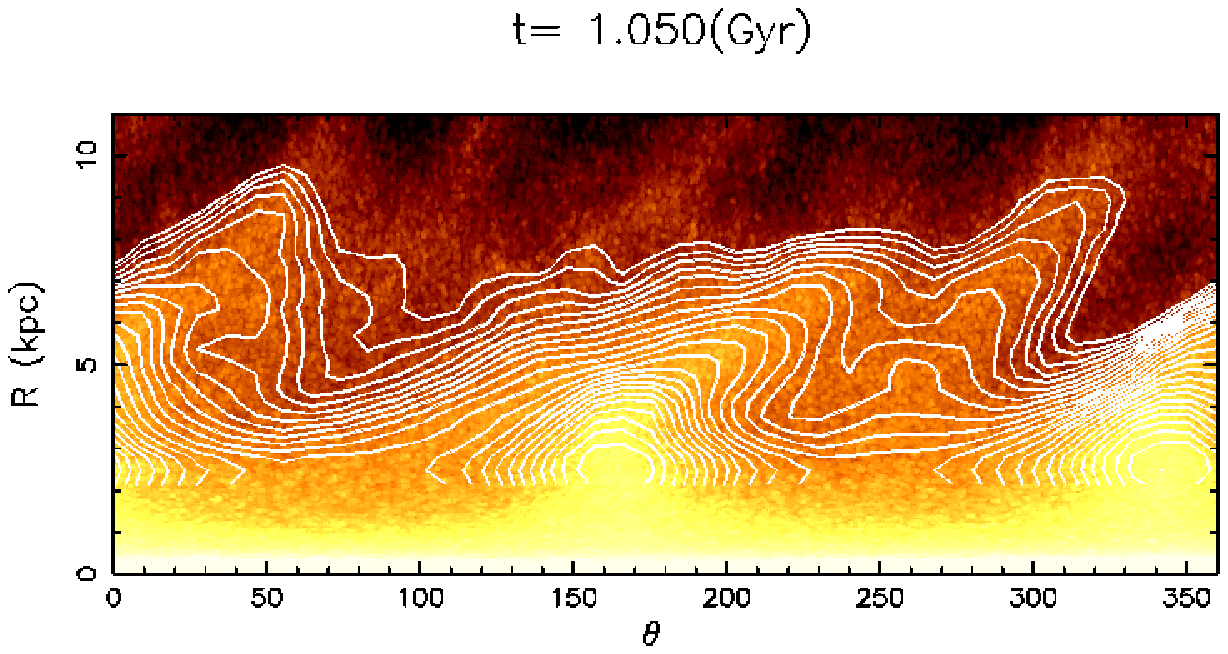}} \\

\caption[]{The density distribution plotted in polar coordinates. Density contours are overplotted in white to identify the highest density regions. The black line that indicates the position of the spiral arm of interest is omitted in the bottom right panel because the double peak at $R \sim 6.5$ kpc presents ambiguity for the density weighting method at this radius.}

\label{contearly}
\end{center}
\end{figure*} 

\begin{figure*}
\begin{center}

  \subfloat{\includegraphics[scale=0.58]{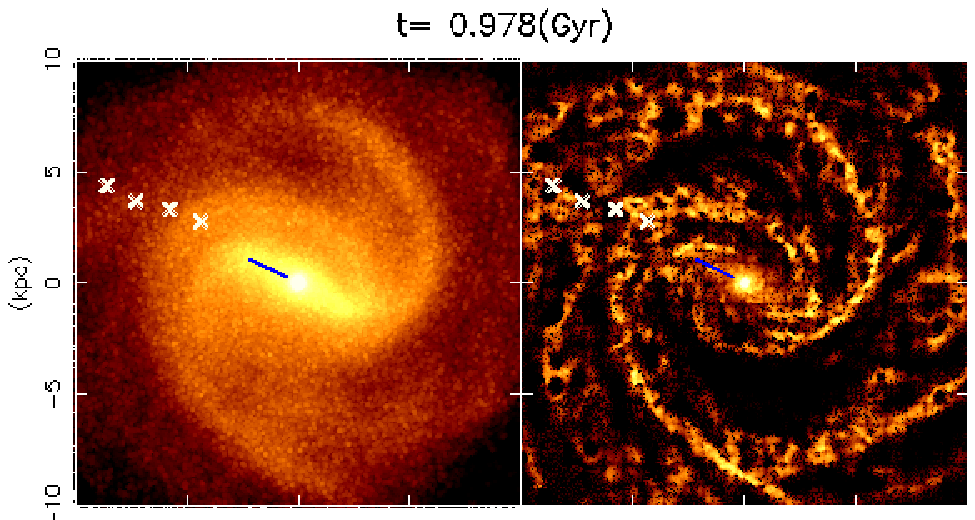}} \hspace{-5.0mm}
  \subfloat{\includegraphics[scale=0.58] {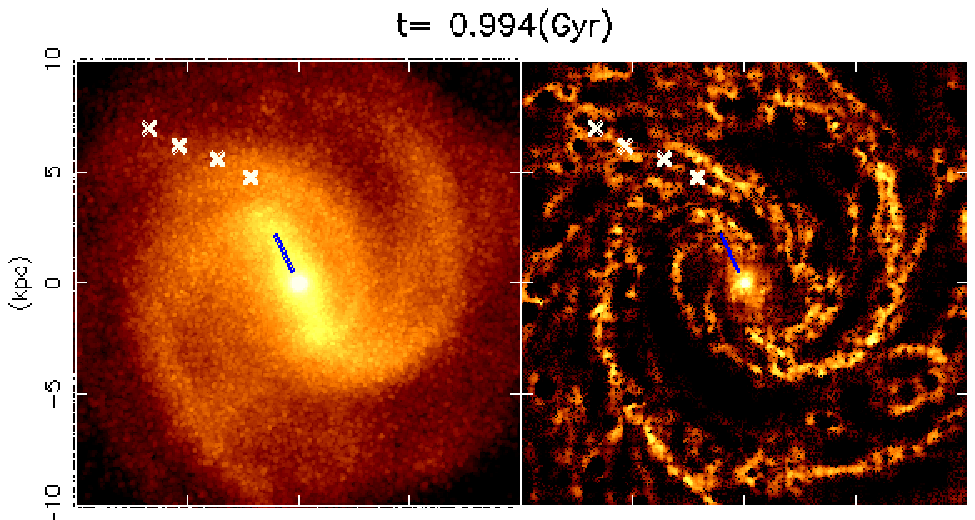}} \hspace{-5.0mm}
  \subfloat{\includegraphics[scale=0.58] {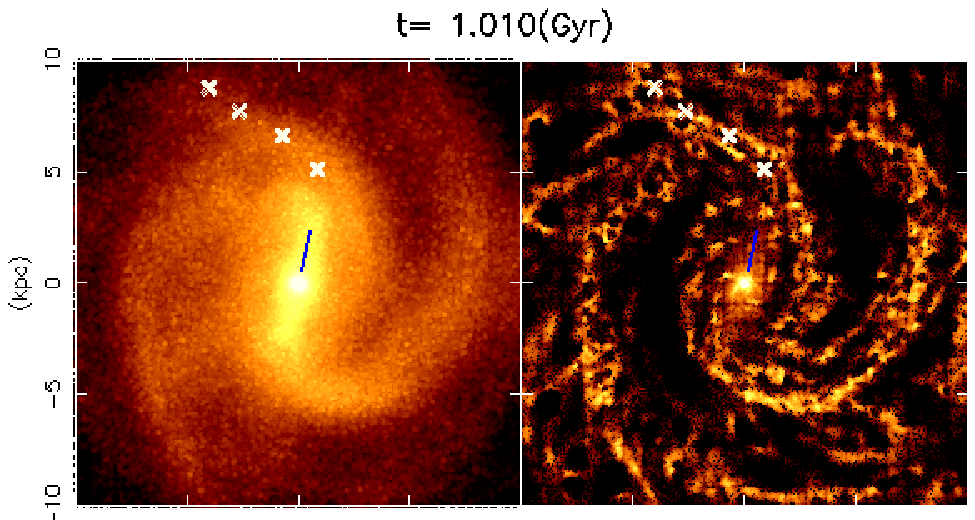}} \\
  \subfloat{\includegraphics[scale=0.58]{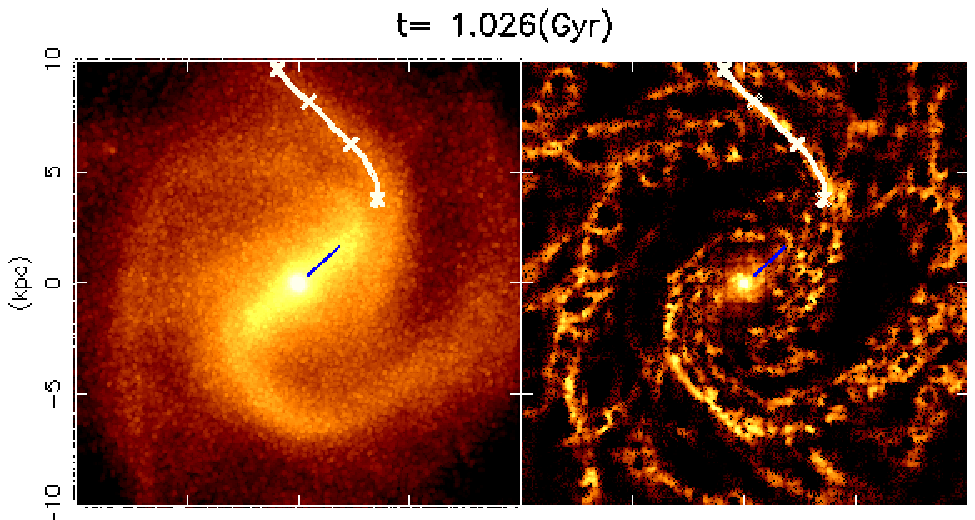}} \hspace{-5.0mm}
  \subfloat{\includegraphics[scale=0.58] {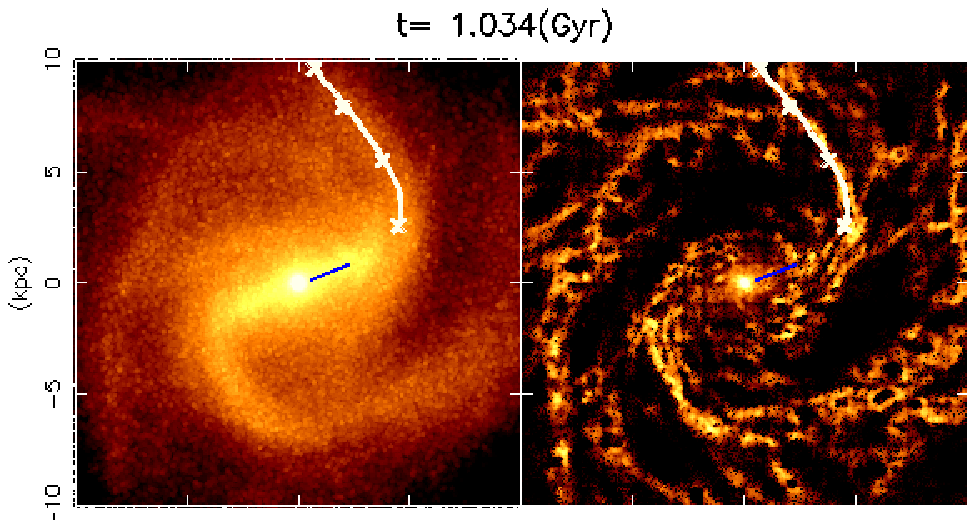}} \hspace{-5.0mm}
  \subfloat{\includegraphics[scale=0.58] {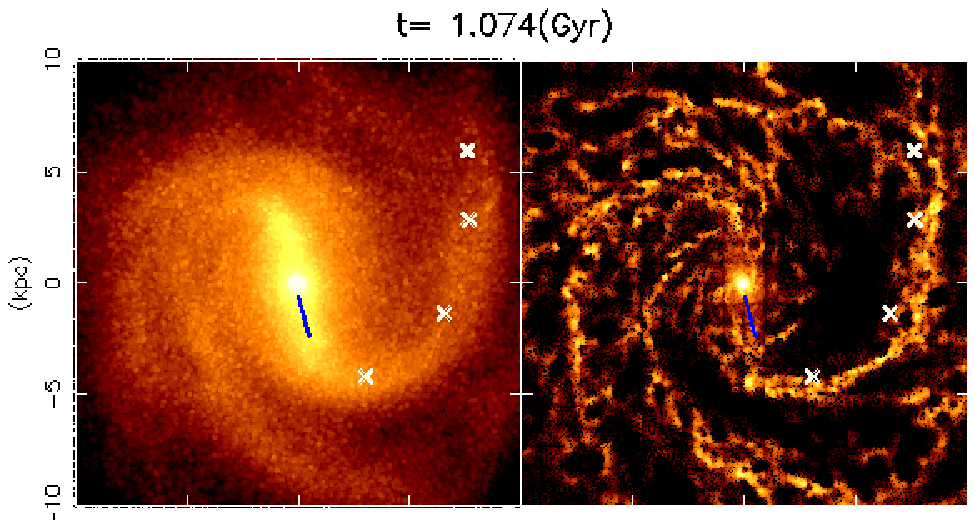}} \\
  \subfloat{\includegraphics[scale=0.58]{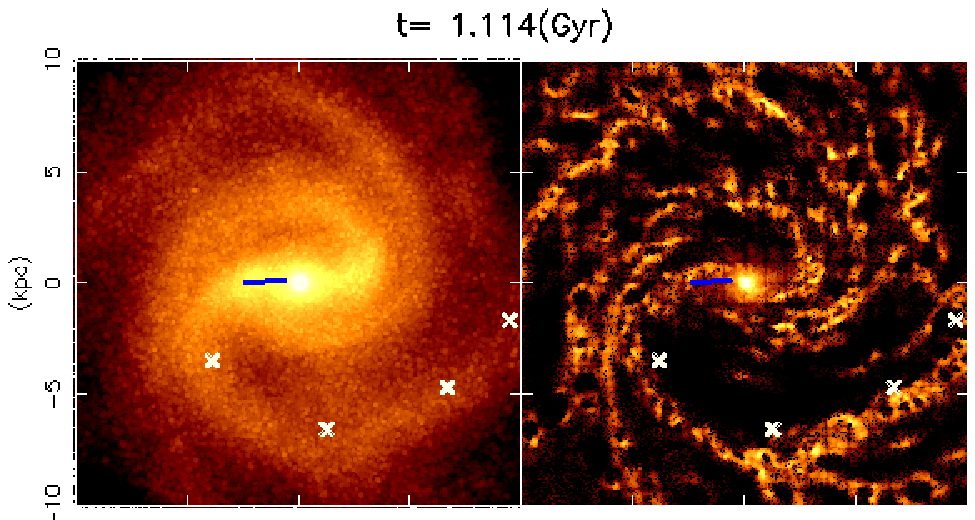}} \hspace{-5.0mm}
  \subfloat{\includegraphics[scale=0.58] {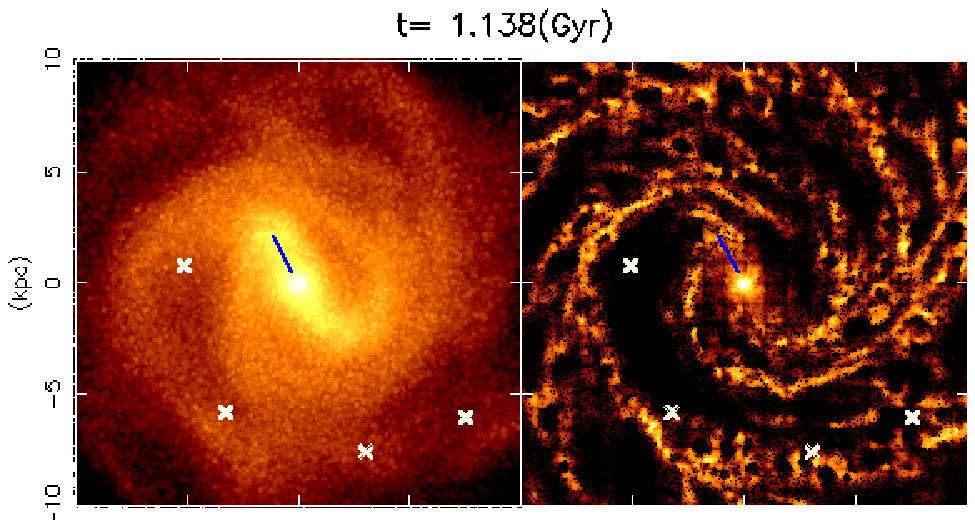}} \hspace{-5.0mm}
  \subfloat{\includegraphics[scale=0.58] {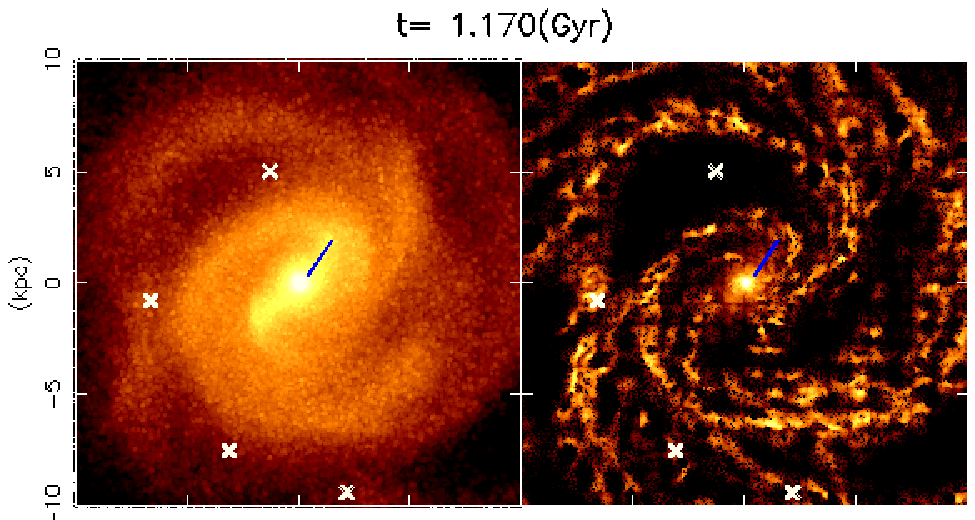}} \\

\caption[]{Snapshots of the stellar and gas disc from $t = 0.978$ to $t = 1.170$ Myr. The blue lines mark out the bar and extend from $1 - 3$ kpc. The white lines mark out the position of highest density over the spiral arm found from the method described in the text,  and extend from $5 -10$ kpc. The spiral arm lines are shown at the centre and middle-left panels only, because the spiral arm in all other snapshots shown here has either not fully formed or displays double peak structures, and could not be fitted well by our method. Anchors are plotted over the spiral arm, and are rotated from the centre snapshot with the rotational velocity.}
\label{dentraceearly}
\end{center}
\end{figure*}

\begin{figure*}
\begin{center}

  \subfloat{\includegraphics[scale=0.65] {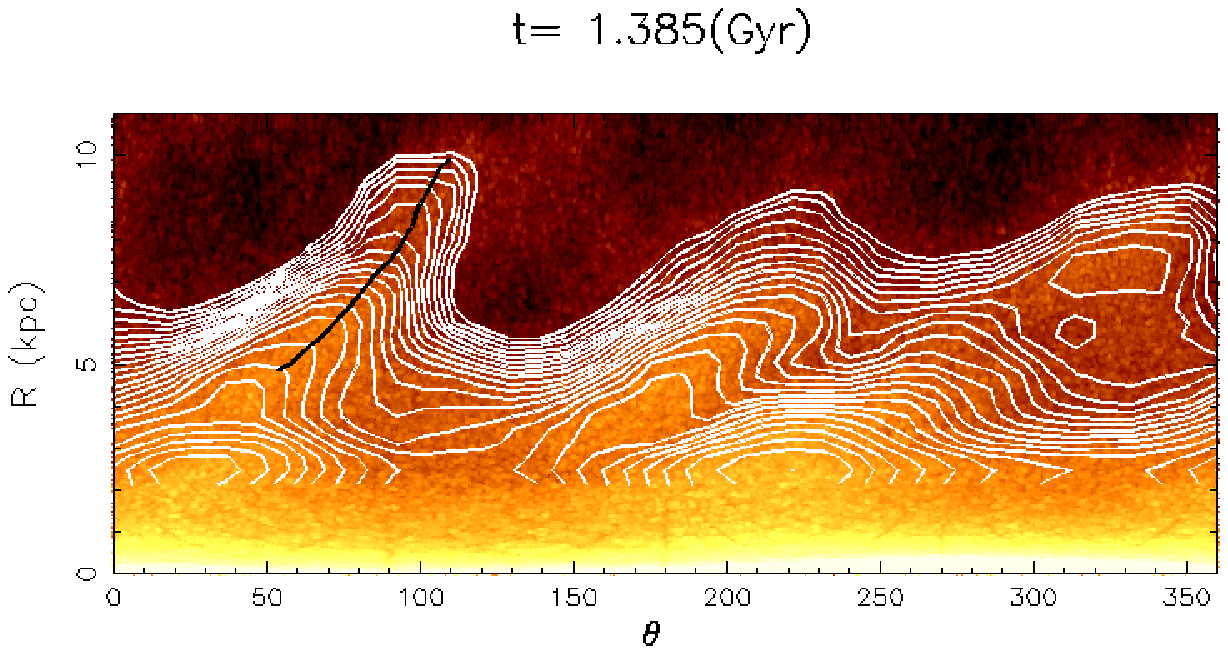}} \hspace{-15.0mm}
  \subfloat{\includegraphics[scale=0.65] {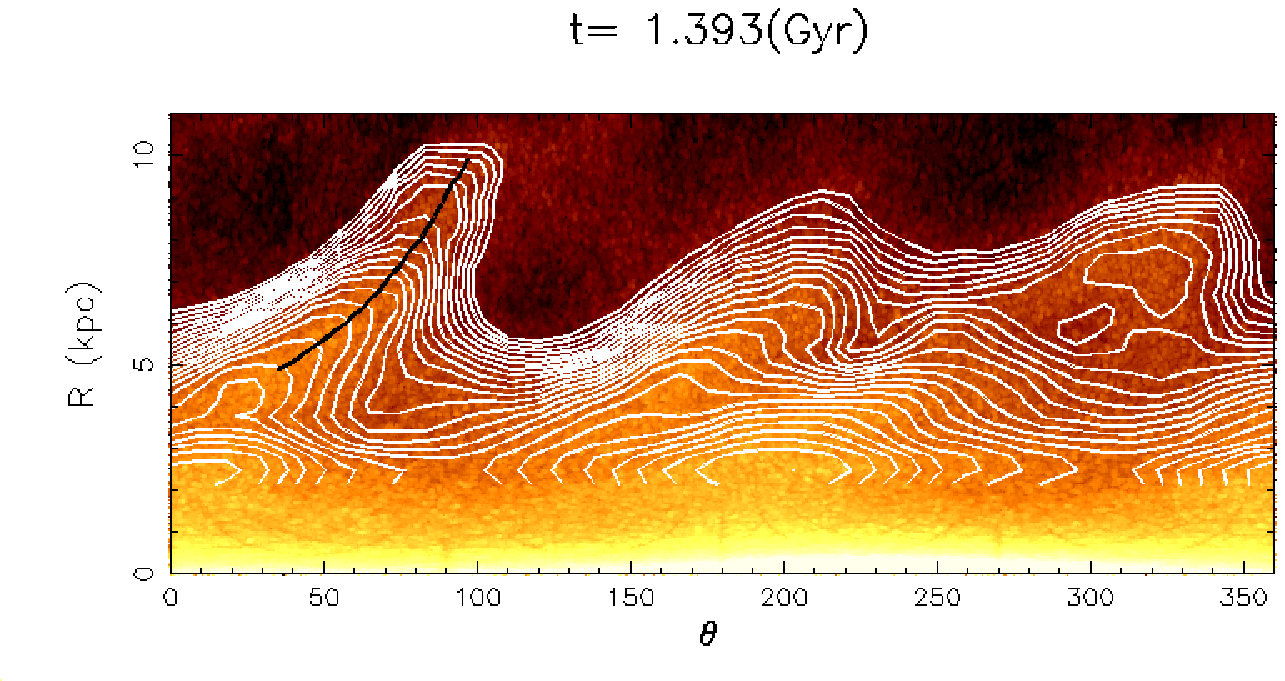}} \\ \vspace{-10.5mm}
  \subfloat{\includegraphics[scale=0.65] {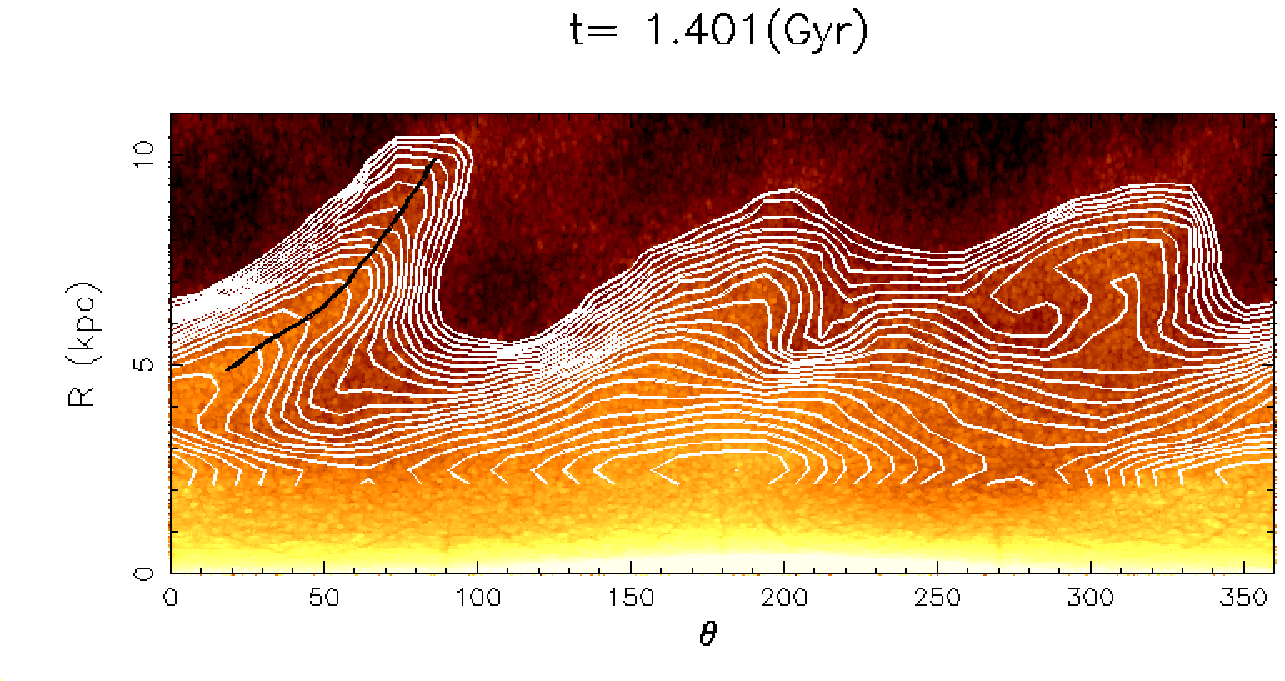}} \hspace{-15.0mm}
  \subfloat{\includegraphics[scale=0.65] {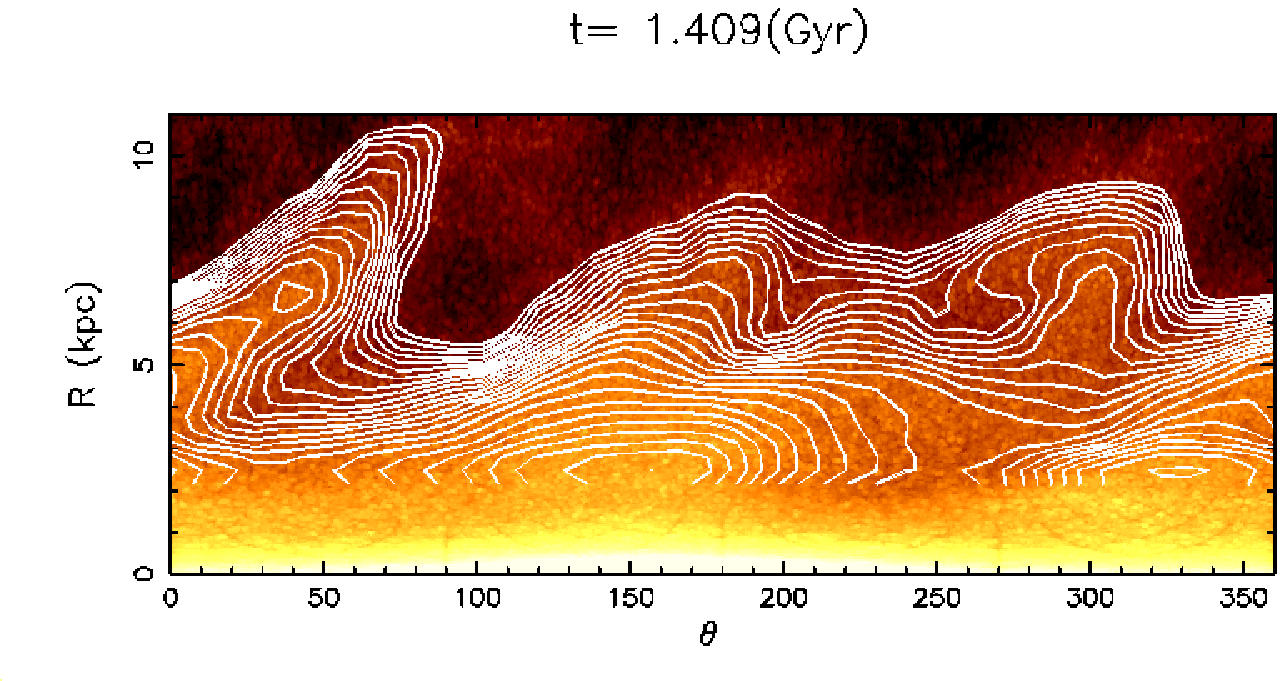}} \\

\caption[]{Same as Fig. \ref{contearly}, but for the late epoch (weak bar case). The black line that highlights the locus of the spiral arm is omitted in the bottom right panel owing to the double peak structure at $R \sim 6.5$ kpc.}
\label{contlate}
\end{center}
\end{figure*} 

\begin{figure*}
\begin{center}

  \subfloat{\includegraphics[scale=0.58]{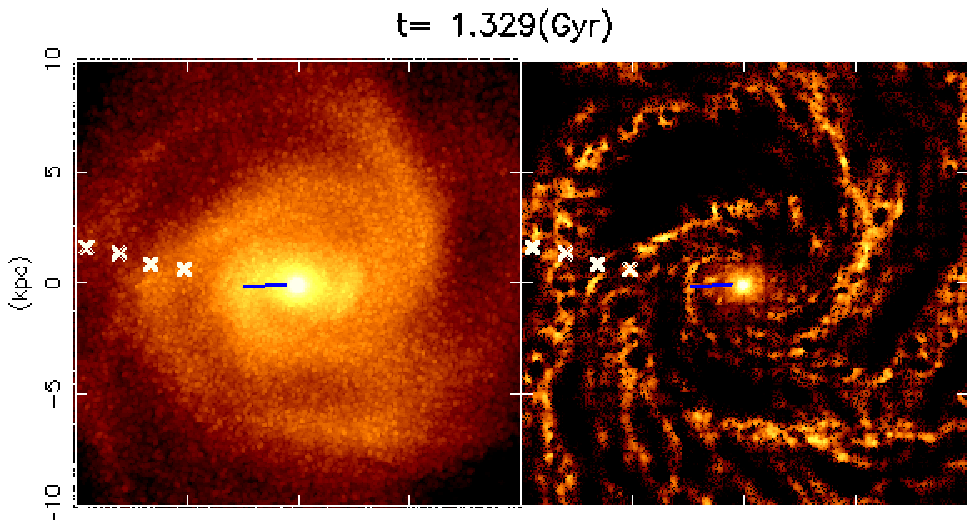}} \hspace{-5.0mm}
  \subfloat{\includegraphics[scale=0.58] {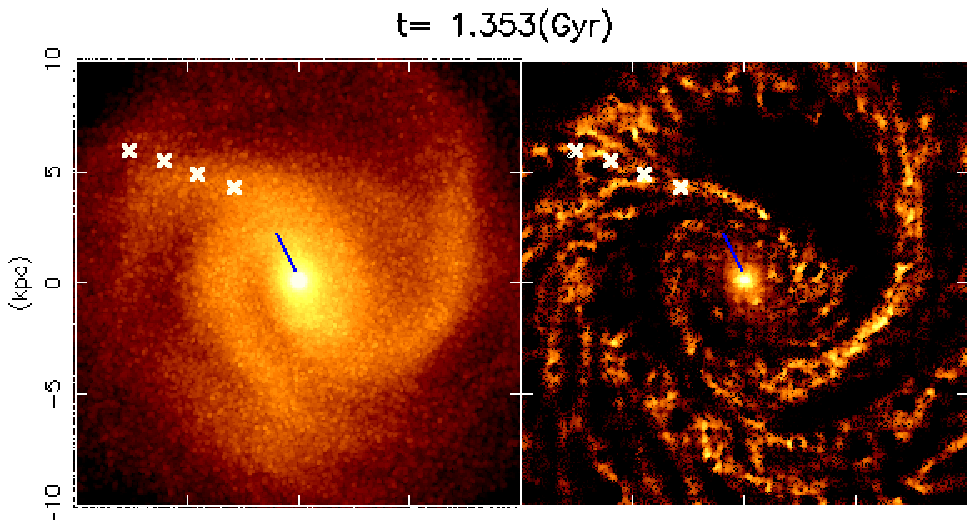}} \hspace{-5.0mm}
  \subfloat{\includegraphics[scale=0.58] {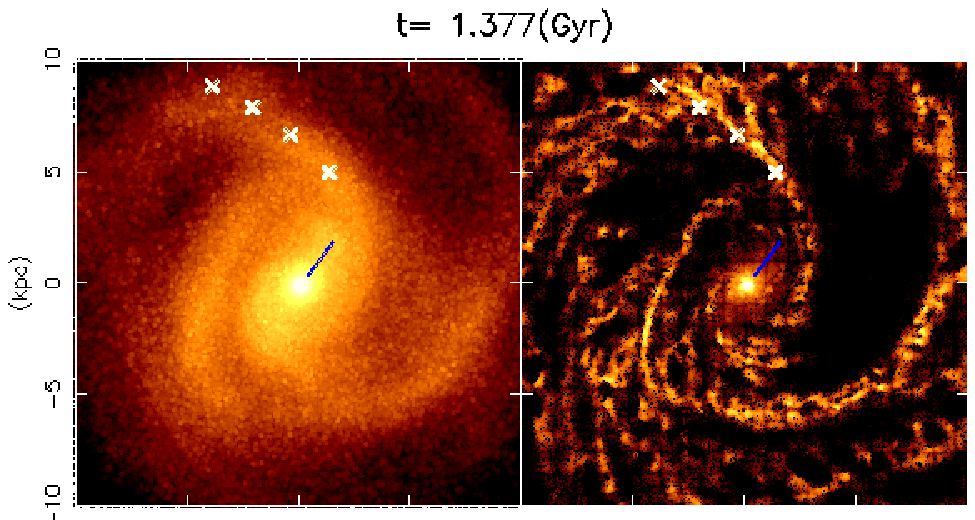}} \\
  \subfloat{\includegraphics[scale=0.58]{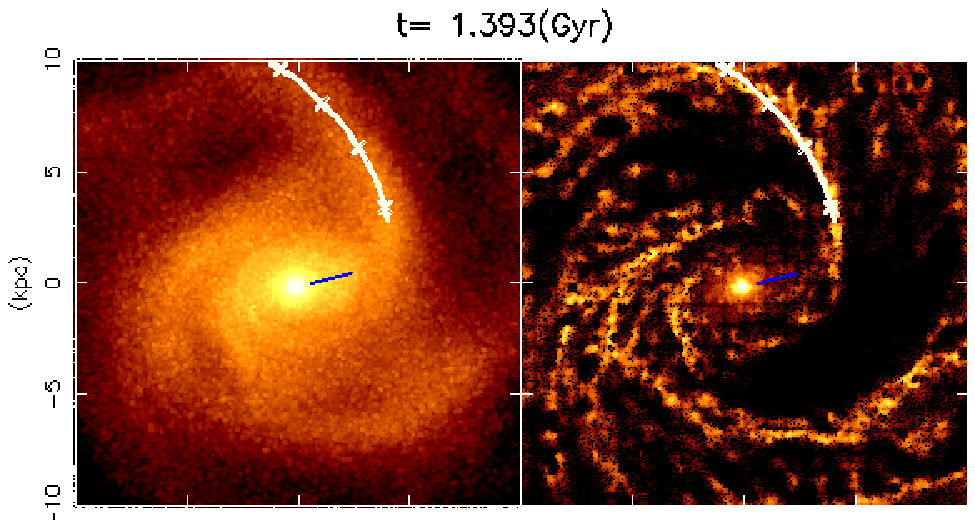}} \hspace{-5.0mm}
  \subfloat{\includegraphics[scale=0.58] {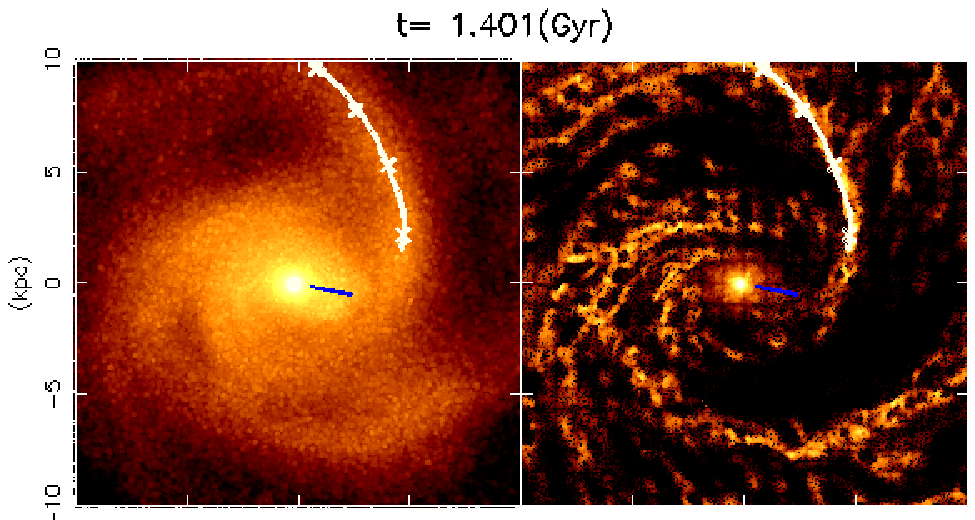}} \hspace{-5.0mm}
  \subfloat{\includegraphics[scale=0.58] {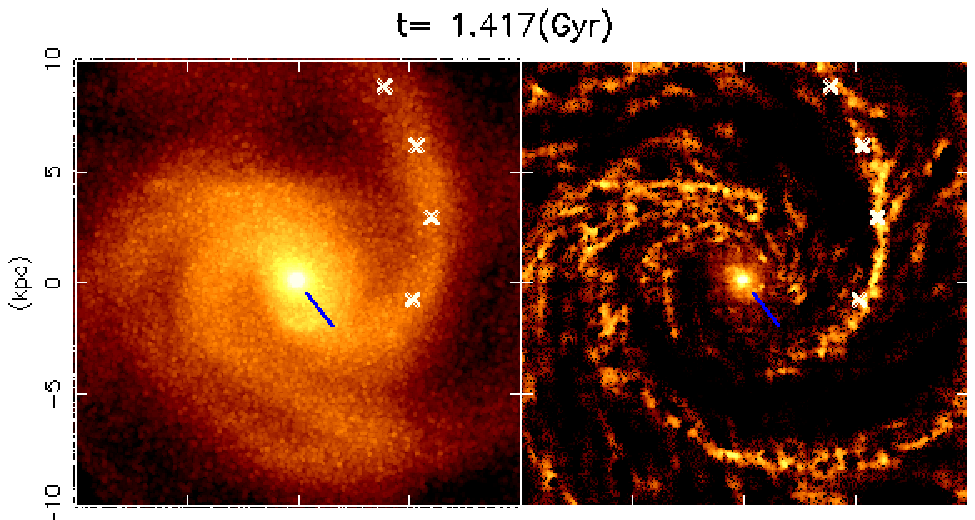}} \\
  \subfloat{\includegraphics[scale=0.58]{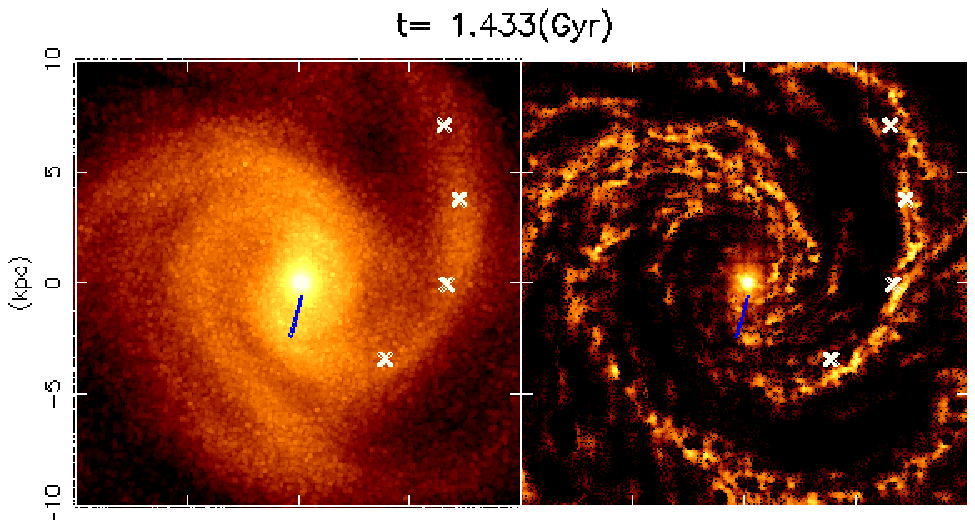}} \hspace{-5.0mm}
  \subfloat{\includegraphics[scale=0.58] {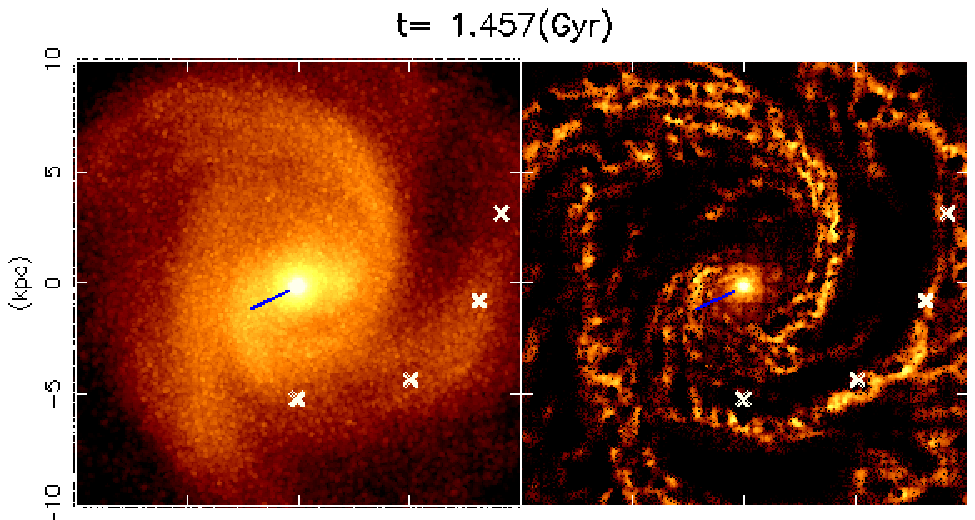}} \hspace{-5.0mm}
  \subfloat{\includegraphics[scale=0.58] {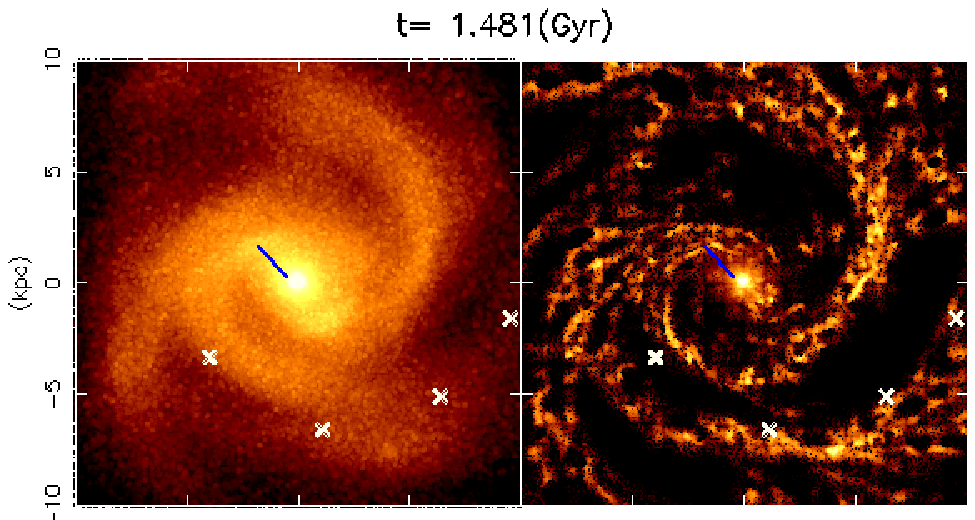}} \\

\caption[]{Same as Fig. \ref{dentraceearly}, but for the late epoch (weak bar case).}
\label{dentrace}
\end{center}
\end{figure*}

\subsection{Pattern Speed}

Here we present a method for calculating the rate at which the stellar density enhancement rotates as a function of radius i.e. the pattern speed. The pattern speeds are often measured by spectrogram analysis (e.g. \citealt{QDBM10}). However, we focus on the angular pattern speed of the apparent spiral feature, and in this paper we refer to this as the pattern speed. The location of the stellar density peak is found at a range of radii for a series of snapshots. This is done by weighting the positional information of particles close to the arm by their density. First, an azimuth coordinate is chosen close to the peak as an initial guess at a given radius. Then, a suitable azimuthal range centred on the initial guess is applied to select the particles covering the whole spiral arm or bar. From the selected particles at a given radius, we calculate:

\begin{equation}
\overline{\theta} _{sp} (r) = \frac{\sum_{i}^{N} \rho _i \theta_i (r)}{\sum_{i}^{N} \rho _i (r)}. 
\label{pkmeth}
\end{equation} 
Here, $\theta _i$ and $\rho _i$ are the azimuth angle and stellar density at the position of the $i$-th star particle. We iteratively find $\overline{\theta} _{sp} (r)$, and narrow the sampling range of $\theta$ progressively. In order to check the reliability of this method and the suitability of spiral arms, we show the density contours plotted over the density map in Fig. \ref{contearly} and Fig. \ref{contlate} at the early and late epoch respectively. The contours show that at some time steps such as $t=1.026$ and $t = 1.385$ Gyr, the spiral arm of interest (($R (\rm{kpc}), \theta (\rm{deg})) \sim (5, 50)$ in both Fig. \ref{contearly} and \ref{contlate}) has a well-defined single peak, which is more suitable for tracing unambiguously. However, at time $t = 1.050$ Gyr ($t = 1.409$ Gyr) in Fig. \ref{contearly} (Fig. \ref{contlate}), the arm develops two peaks at a radius around $6.5$ kpc as it begins to break. Therefore, to remain robust, the peak tracing method is restricted to those snapshots where the azimuthal density distribution around the spiral arm is made of a single peak at each radius. The bar is unaffected by this caveat, and is traced at many snapshots.

\begin{figure}
\centering
\includegraphics[scale=0.42]{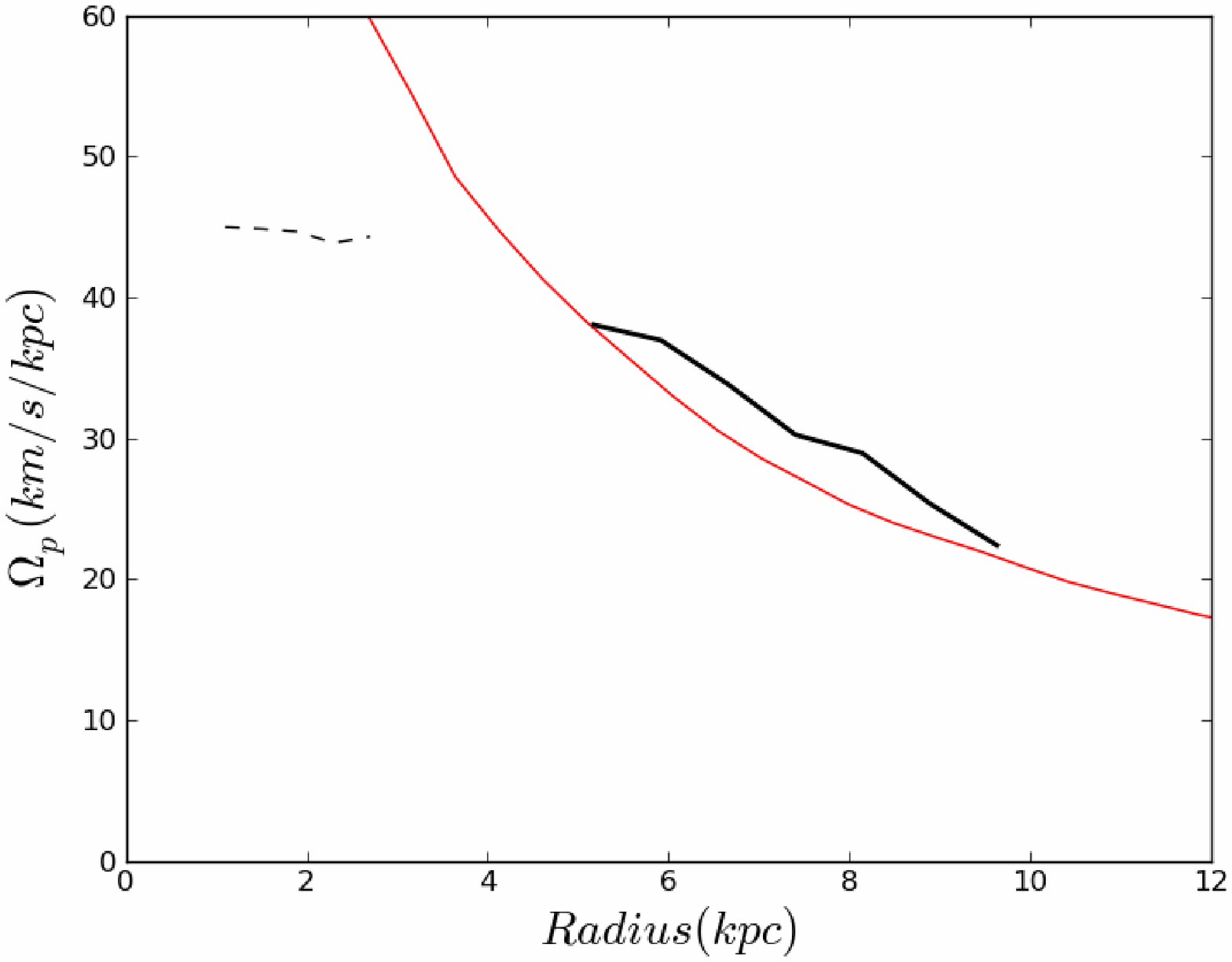}
\caption[omgp_vrot_2591]
{The bar and spiral arm pattern speed calculated for the snapshots shown in Fig. \ref{dentraceearly}. The spiral arm pattern speed (solid black line) is averaged over several pattern speeds calculated at different snapshots over the course of the spiral arm's evolution. The rotational velocity is also plotted (solid red line). The bar pattern speed (dashed black line) is found to be $\sim 45$ km $\rm s^{-1}$ $\rm  kpc^{-1}$. The spiral pattern speed exhibits a decreasing trend with radius that is similar to but slightly faster than the rotational velocity.}
\label{omgpearly}
\end{figure}

\begin{figure}
\centering
\includegraphics[scale=0.42]{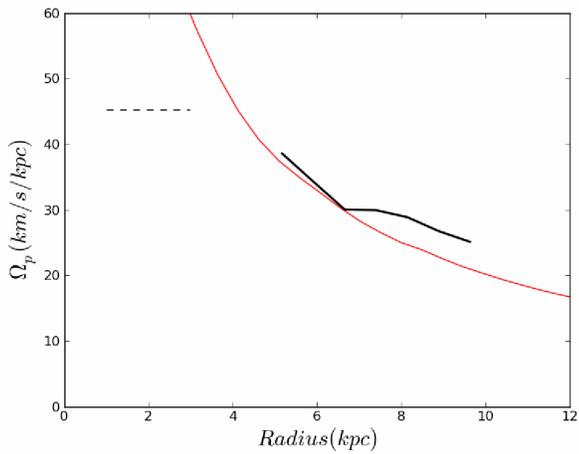}
\caption[omgp_vrot_2591]
{The same as Fig. \ref{omgpearly}, but got the pattern speed of the spiral arm of the late epoch when the bar is relatively weak.}
\label{omgp}
\end{figure}

\begin{figure}
\centering
\includegraphics[scale=0.42]{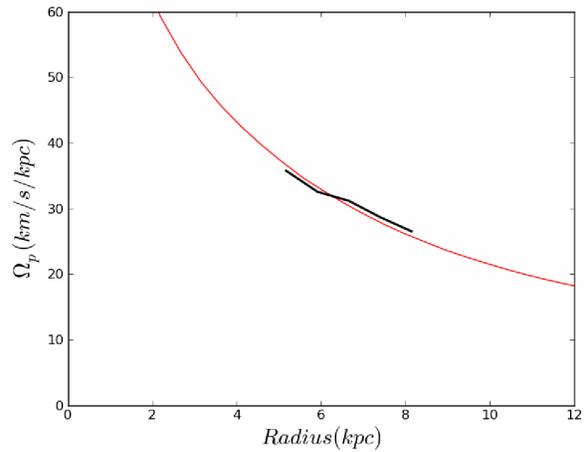}
\caption[omgp_vrot_2591]
{The spiral arm pattern speed of the simulated galaxy presented in \citet{GKC11} (black line) and the rotational velocity at the corresponding time (red line). This galaxy has no bar.}
\label{omgpdm}
\end{figure}

\begin{figure}
\begin{center}
\hspace{-8.5mm}
\includegraphics[scale=0.3]{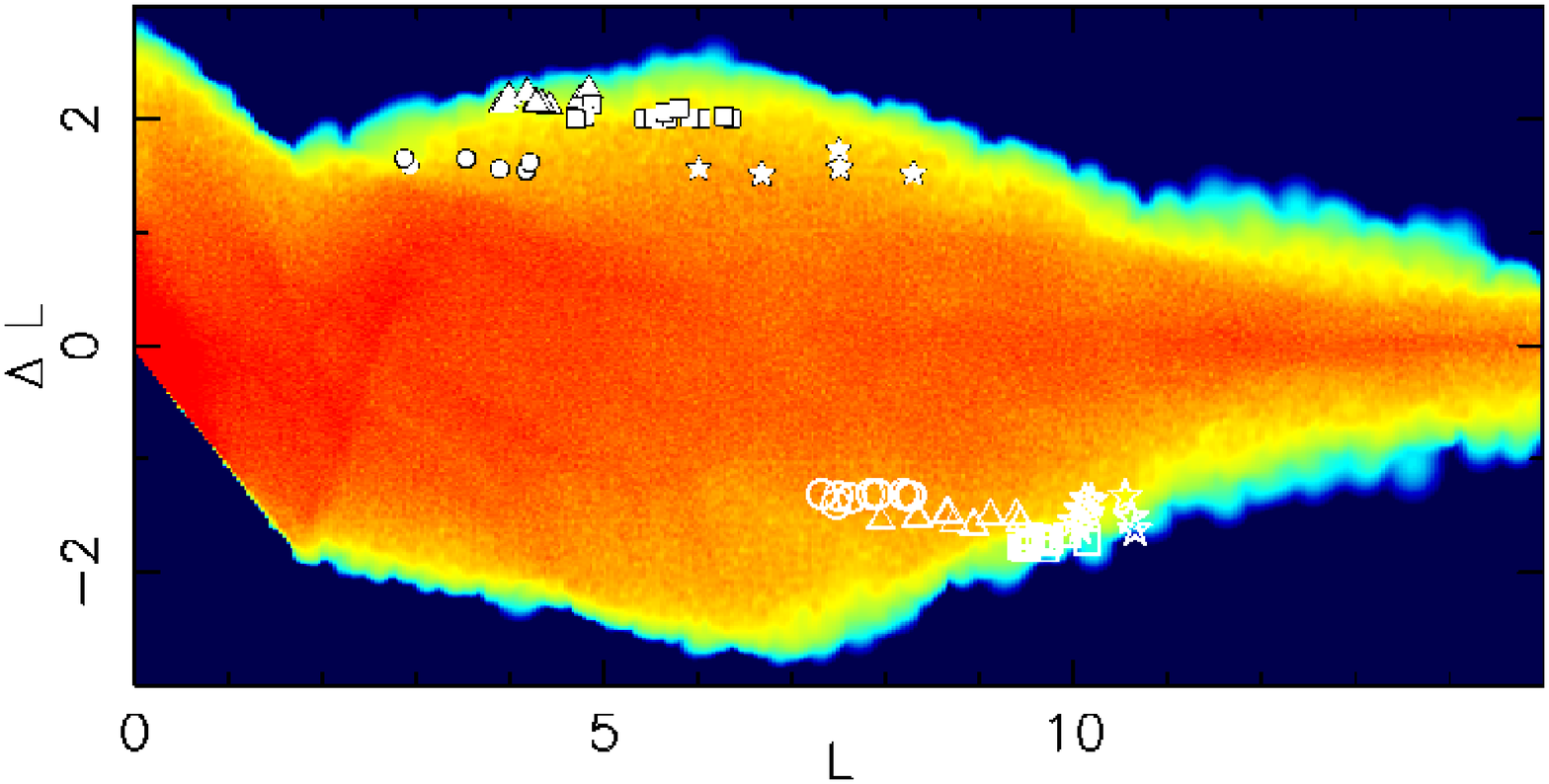}
\caption{The angular momentum, $L$, at the beginning of the late epoch time period (40 Myr before $t=1.393$ Gyr) plotted against the change in angular momentum, $\Delta L$, $80$ Myr later. The smoothed colour map from low number density (dark blue) to high number density (red) incorporates all disc star particles, and shows a broad angular momentum range for migration. Selected positive (filled symbols) and negative (open symbols) extreme migrators (see text) are highlighted by circle (chosen at a radius of $\sim 6$ kpc), triangle ($\sim 7$ kpc), square ($\sim 8$ kpc) and star ($\sim 9$ kpc) symbols. Units are arbitrary.}
\label{deltallate}
\end{center}
\end{figure}

The peak density of the spiral arm ($5 < R < 10$ kpc) and the bar ($1 < R < 3$ kpc) is shown in Fig. \ref{dentraceearly} and Fig. \ref{dentrace} for the early and late epoch respectively. For clarity, anchors (marked by crosses) are placed at four radii spread over the above radial range of the spiral arm. Their positions are initially selected at $t=1.034$ and $t=1.393$ Gyr (early and late epoch respectively) according to the spiral arm peak line traced at that time step. Their positions at other time steps are calculated by rotating the anchors with mean rotational velocity at the radius at which the anchors are located. The gas maps (right panels of Figs. \ref{dentraceearly} and \ref{dentrace}) give some indication to how the spiral arm evolves during the formation stage. For example, in Fig. \ref{dentrace}, at $t = 1.353$ Gyr, there appear to be two gas arms in the outer radii, one of which is marked by the anchor points. This arm seems to originate from a previously wound up arm that merges with another arm to form the spiral arm for which we trace the peak density at $t = 1.393$ Gyr. The anchors plotted on the star and gas maps clearly show that the apparent spiral arm follows a shearing pattern speed close to the mean rotation of star particles, and helps to define the time of formation, $t_f \sim 0.994$ Gyr and destruction, $t_d \sim 1.170$ Gyr for the early epoch. This gives a lifetime of $\tau \sim 180$ Myr. For the late epoch, the formation time, $t_f \sim 1.353$ Gyr and destruction time, $t_d \sim 1.481$ Gyr, give a lifetime of $\tau \sim 130$ Myr.

We calculate the pattern speed at the early epoch, when the bar is relatively strong. This is shown in Fig. \ref{omgpearly}, which appears to be decreasing with radius, but is faster than the rotational velocity. For comparison with the relatively weak bar case, we calculate the pattern speed at the late epoch, which is shown in Fig. \ref{omgp}. The pattern speed appears to be similar to the mean rotation of star particles in the inner regions ($5 < R < 7$ kpc), and is faster in the outer regions ($7 < R < 10$ kpc). Aside from the kink at $R \sim 6.5-7$ kpc, the pattern speed again appears to decrease with the radius.  The flattening at $R \sim 6.5-7$ kpc is approximately the same radius at which a break is observed in the density contours over plotted in Fig. \ref{contlate}. The pattern speed of this late epoch is slower than the pattern speed of the earlier epoch when the bar is stronger. This indicates that the bar may boost the pattern speed somewhat, and cause it to become slightly faster than rotational velocity. Nevertheless, the anchors in Fig. \ref{dentraceearly} show that the spiral arm rotates in a similar way to the star particles at both epochs.

Further comparison is made with the galaxy presented in \citet{GKC11}, which has no bar or gas component. We apply the same peak tracing technique in equation (\ref{pkmeth}), and find the pattern speed of arms that have suitable single peaks. This is plotted in Fig. \ref{omgpdm}. The calculated pattern speed follows the same trend as reported in \citet{GKC11}, i.e. is the same as rotational velocity at all radii, and therefore the spiral arm co-rotates with the star particles. Therefore, our simulation shows that the transient, winding spiral arms occur in barred galaxy simulations (see also \citealt{BAM09}), but the pattern speed appears to be boosted slightly out of co-rotation by the bar feature. Furthermore, although there are only three cases studied here and the differences are relatively small, comparison between this non-barred case and the two barred cases indicates that the pattern speed becomes faster with increasing bar strength, which deserves further investigation.

\begin{figure*}
\begin{center}

  \subfloat{\includegraphics[scale=0.5]{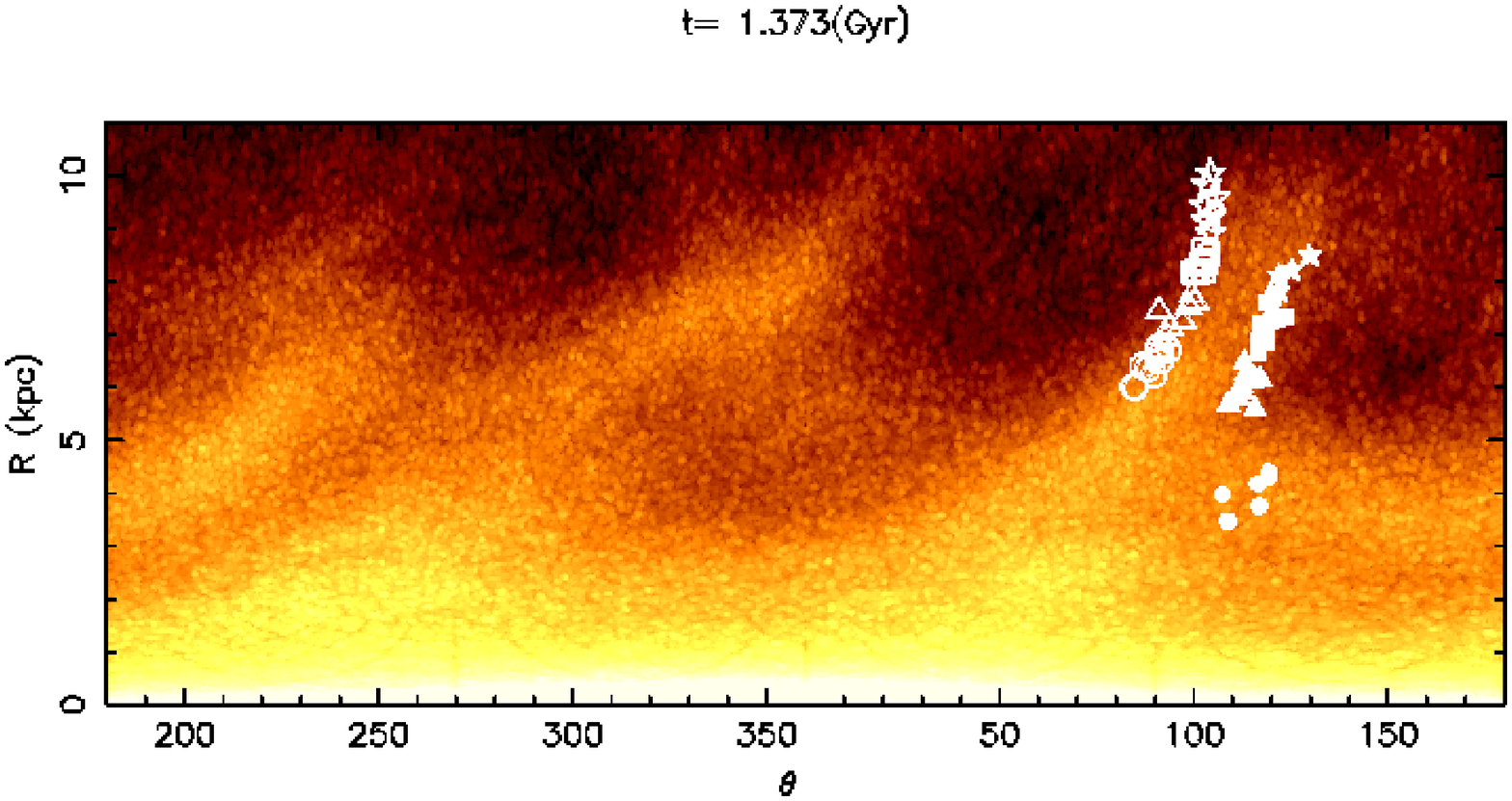}} \vspace{-14.0mm}\\
  \subfloat{\includegraphics[scale=0.5] {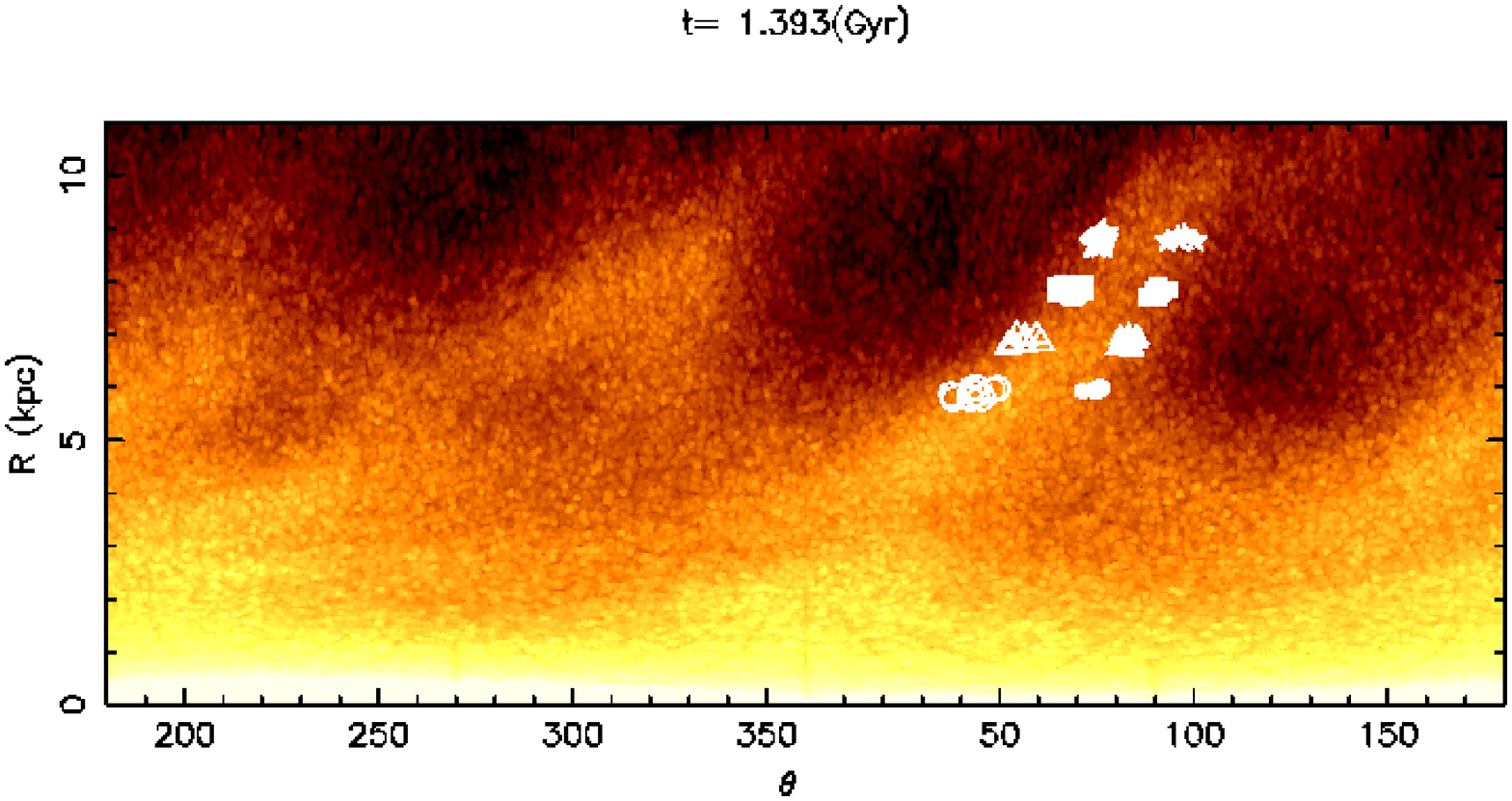}} \vspace{-14.0mm}\\
  \subfloat{\includegraphics[scale=0.5] {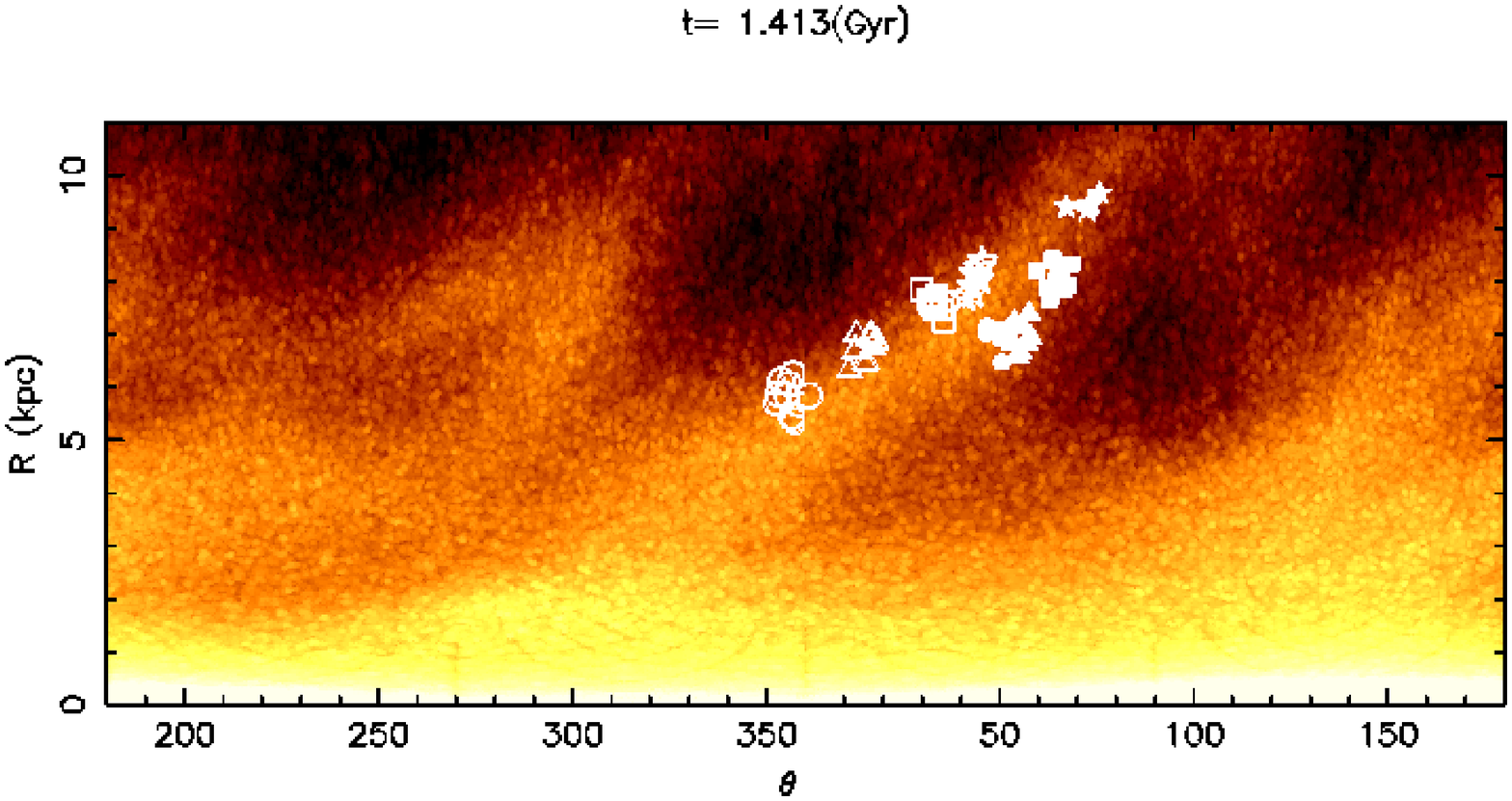}} \\

\caption[]{Stellar density distribution plotted in polar coordinates showing the time evolution of four sets of extreme migrators selected from particle samples around radii of 6, 7, 8 and 9 kpc at the $t = 1.393$ Gyr snapshot. The symbols correspond to those of Fig. \ref{deltallate}. Azimuth ($\theta$) is expressed in degrees. The particles tend to migrate toward outer radii on the trailing edge of the spiral arm (filled symbols) and inwards on the leading edge (open symbols).}

\label{5kpcplots}
\end{center}
\end{figure*}

\subsection{Radial Migration around the spiral arm}

Radial migration at the co-rotation radius has been predicted to preserve the circular motion of orbits i.e. not to heat them kinematically (e.g. \citealt{LBK72}; \citealt{SB02}; \citealt{R08}; \citealt{SBl09}; \citealt{RD11}; \citealt{MF10}; \citealt{MFC11}; \citealt{BCP11}). \citet{GKC11} find that the spiral arms are co-rotating at all radii in a non-barred pure N-body disc. As a result, radial migration occurs over a wide range of radii. The spiral arms focused on in this paper rotate slightly faster than the mean rotation velocity. We examine if radial migration still occurs at all radii.

Because the results of the early epoch are the same as the late epoch, we present the particle motion from the late epoch only. First, we select a sample of particles around a given radius of our chosen arm at the $t=1.393$ Gyr snapshot in Fig. \ref{galprev}. In order to focus on star particles that are likely affected by the spiral arm, the particle sample is selected to be within the vertical height of $|z|<200$ pc and azimuthal width of 4 kpc centred on highest density point of the arm i.e. a given radius of the peak line shown in Fig. \ref{dentraceearly}. The radial thickness of the sample is chosen to be 0.25 kpc to ensure that a large sample of stars of approximately the same radius are chosen. 

From the selected sample of particles, we compute the angular momentum change, $\Delta L$, over a period of 80 Myrs and choose those that exhibit the largest values of $| \Delta L |$, some of which are highlighted by symbols in Fig. \ref{deltallate}. As a fraction of their initial angular momentum, $L$, this can be up to as much as $| \Delta L / L | \simeq 50\%$. Note that the angular momentum exchanges in this simulation are much stronger than those of \citet{GKC11}, probably because the spiral arm structure is much more prominent in this barred spiral galaxy. The radii of the guiding centres of these high $|\Delta L|$ particles highlighted in Fig. \ref{deltallate} change significantly i.e. they migrate radially. We call these strongly migrating particles \textquotesingle extreme migrators\textquotesingle , \hspace{0.1mm} and further divide them into two subcategories: positive extreme migrators and negative extreme migrators for particles that gain and lose their angular momentum in the sample respectively. 

In Fig. \ref{5kpcplots} we show three successive snapshots during the migration period of the four extreme migrator samples highlighted in Fig. \ref{deltallate}, each selected around radii of 6, 7, 8 and 9 kpc (positive and negative migrators are denoted by filled and open symbols respectively, where each type of symbol corresponds to a specific selection radius of a sample): $20$ Myr before selection (top panel), at selection (middle panel) and $20$ Myr after selection (bottom panel). The density snapshots for the stellar component are coordinate transformed from cartesian to polar in order to make the radial motion of the selected star particles with respect to the arm clearer. The positive migrators are always seen on the trailing side of the spiral arm and migrate towards the outer radii. They are trapped by the potential of the spiral arm, which accelerates them. During migration to outer radii, instead of passing through the spiral arm they remain in the vicinity of the arm on the trailing side. Therefore, they continue to be accelerated until the spiral arm is disrupted. The negative migrators are particles that migrate towards the inner radii on the leading side of the spiral arm. They are decelerated as they become caught in the potential of the spiral arm, and they continue to decelerate on the leading side, again until the spiral arm is disrupted. This illustrates the different systematic motion that occurs on each side of the spiral arm, which is reminiscent of the behaviour found in \citet{GKC11}. 

\begin{figure}
\begin{center}

  \subfloat{\includegraphics[scale=0.28]{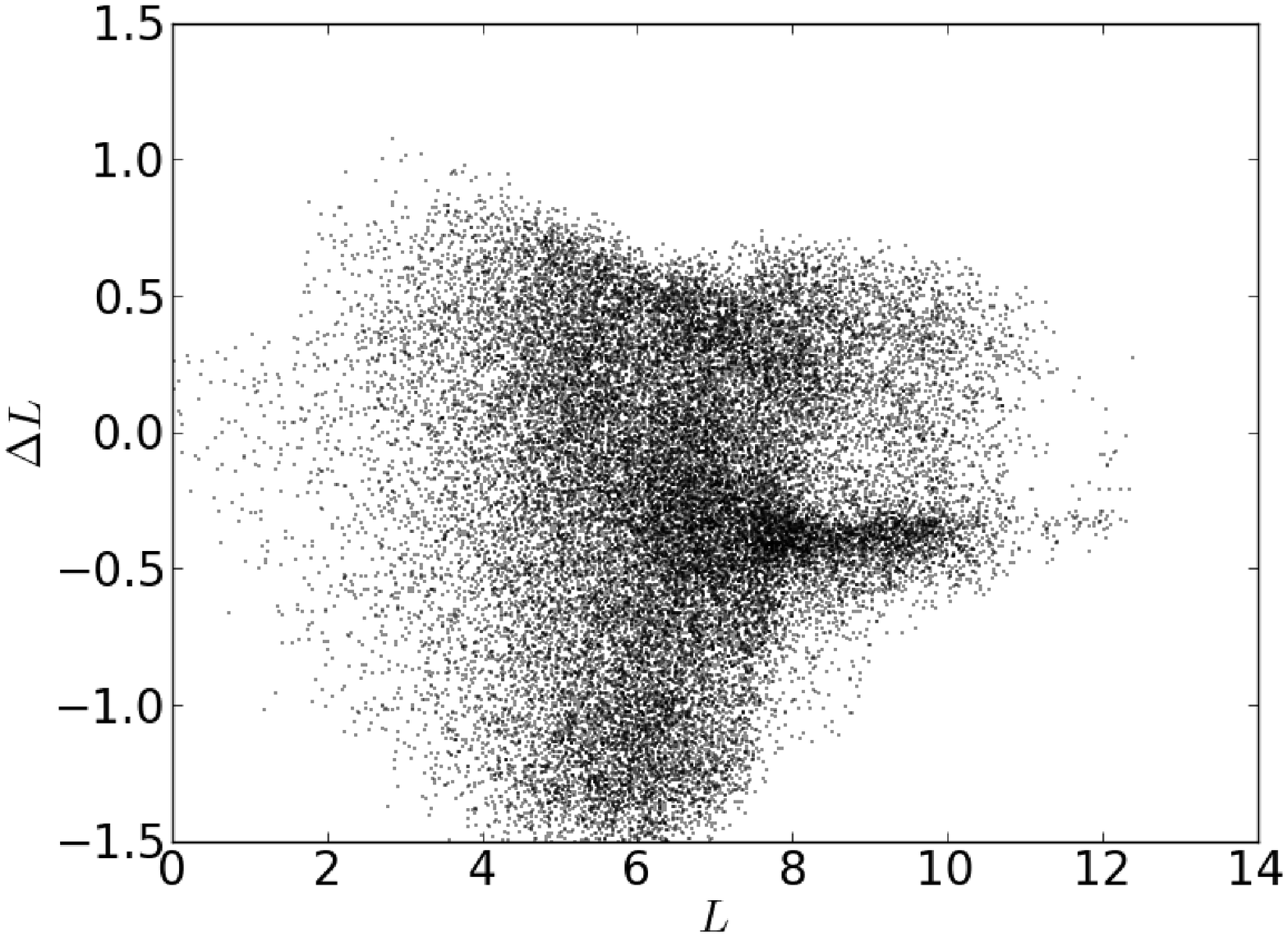}} \vspace{-6.0mm}\\
  \subfloat{\includegraphics[scale=0.28] {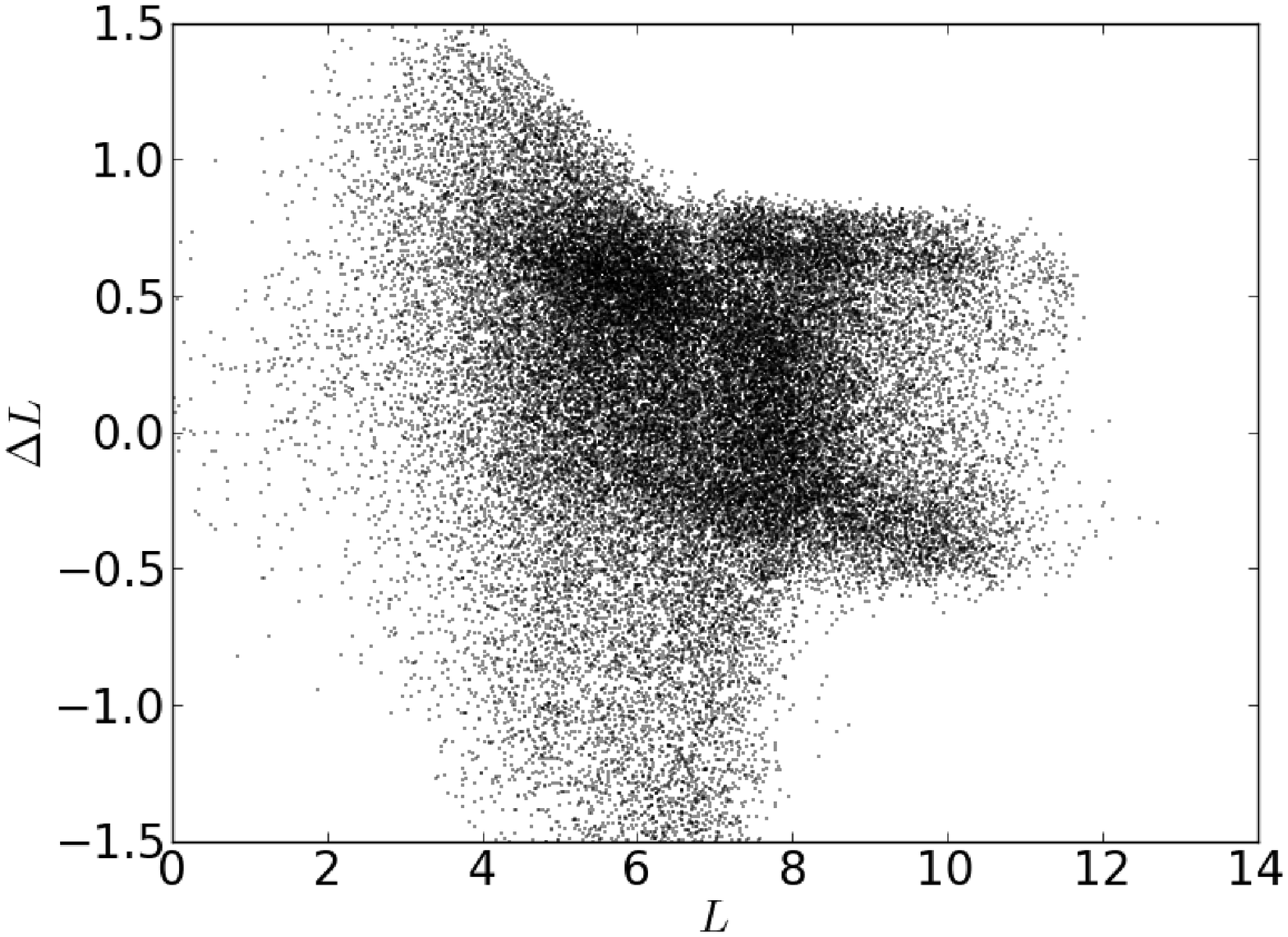}} \vspace{-6.0mm}\\
    \subfloat{\includegraphics[scale=0.28] {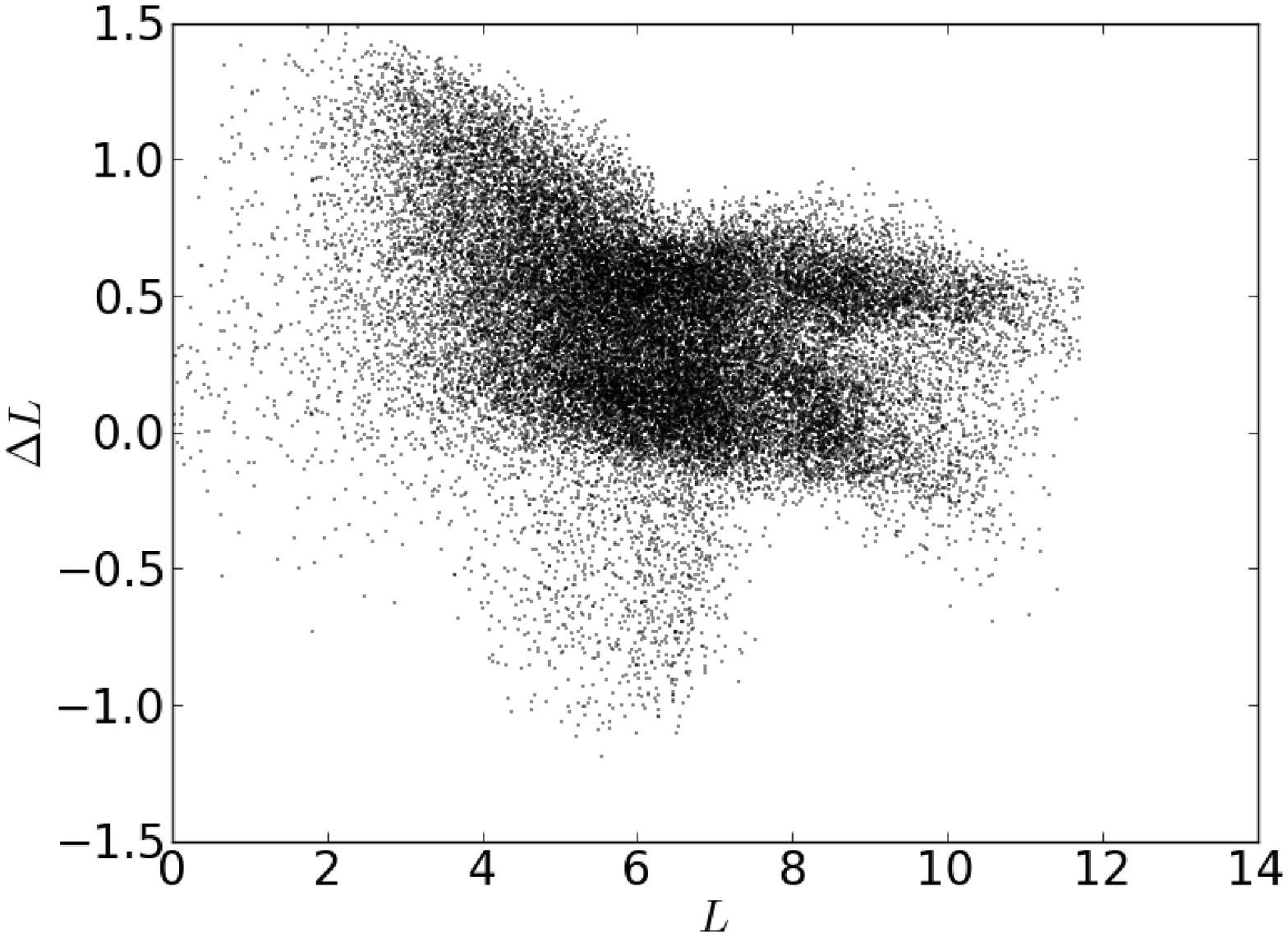}} \vspace{-6.0mm}\\
  \subfloat{\includegraphics[scale=0.28] {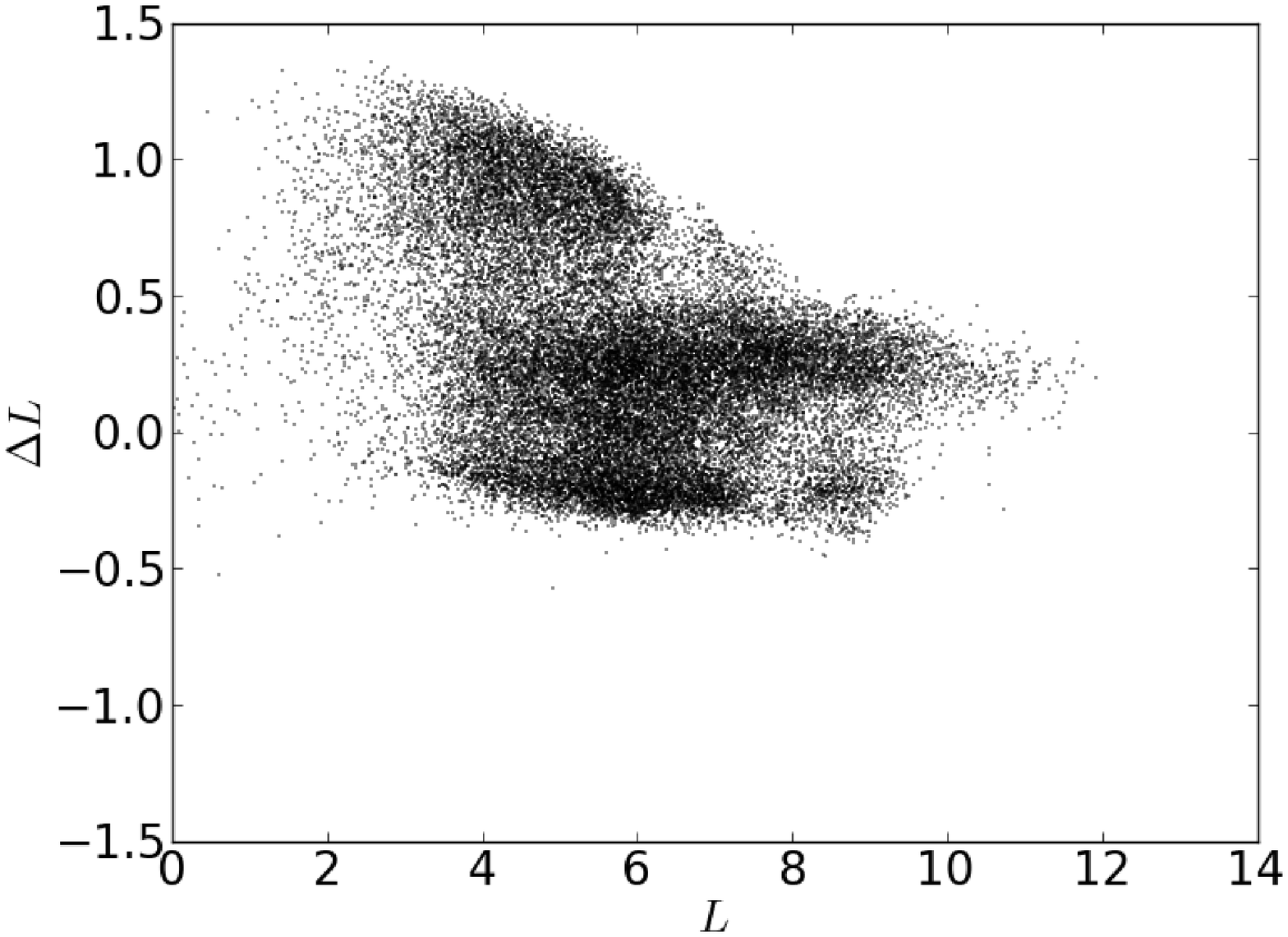}} \\

\caption[]{Initial angular momentum, $L$, as a function of the change in angular momentum, $\Delta L$, within a 40 Myr window for samples of star particles located around the spiral arm of the early epoch. Each panel represents a stage of the spiral arm lifetime. From top to bottom: formation (centred on $t=0.994$ Gyr); fully formed single peak spiral arm ($t=1.034$ Gyr); double peak spiral arm ($t=1.074$ Gyr); breaking ($t=1.114$ Gyr). The strongest migrations occurs at the stage when the arm is fully formed and single peaked (second panel). At later stages, the migration is less.}

\label{LdLearly}
\end{center}
\end{figure} 

\begin{figure}
\begin{center}

  \subfloat{\includegraphics[scale=0.28]{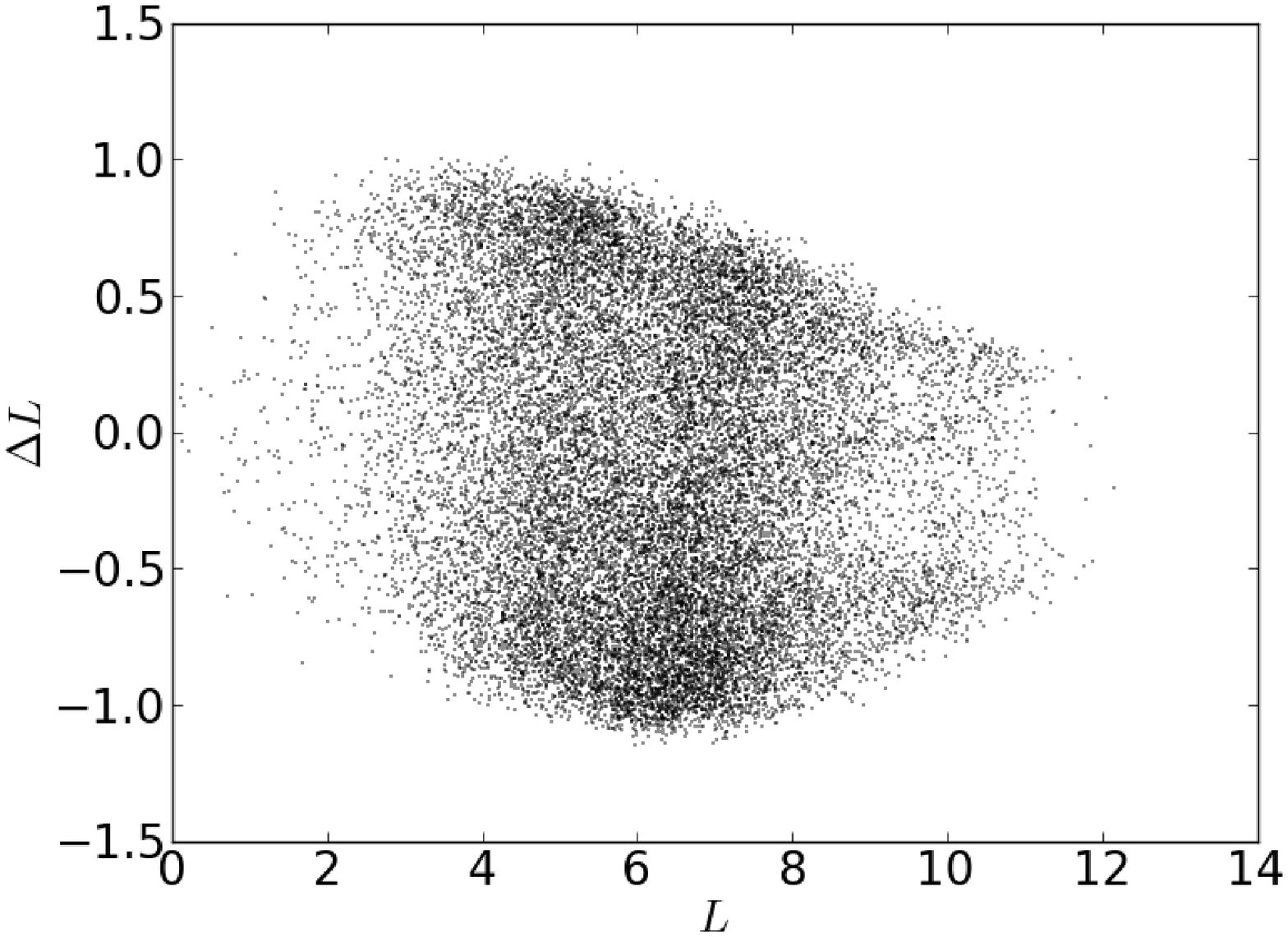}} \vspace{-6.0mm}\\
  \subfloat{\includegraphics[scale=0.28] {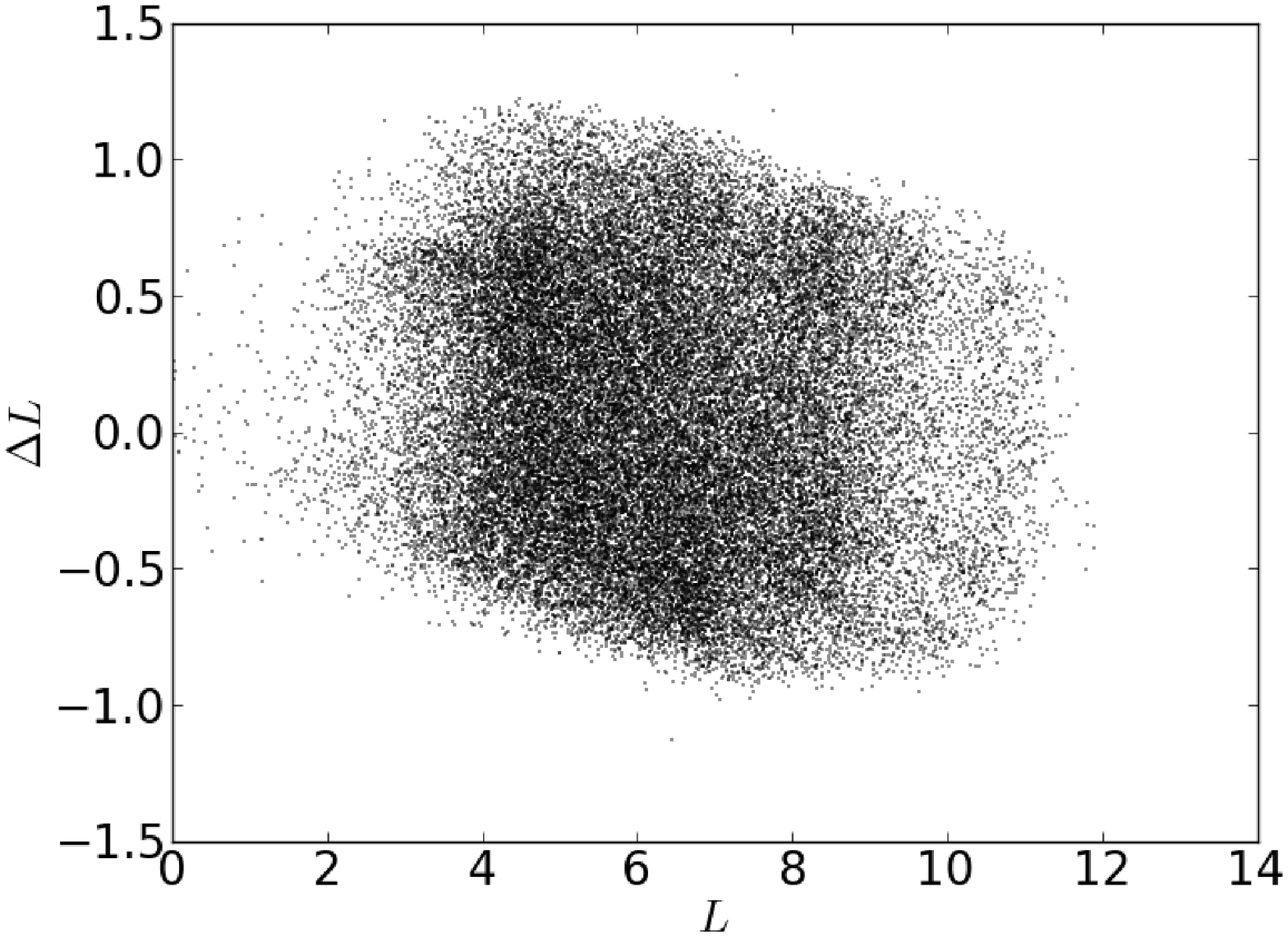}} \vspace{-6.0mm}\\
  \subfloat{\includegraphics[scale=0.28] {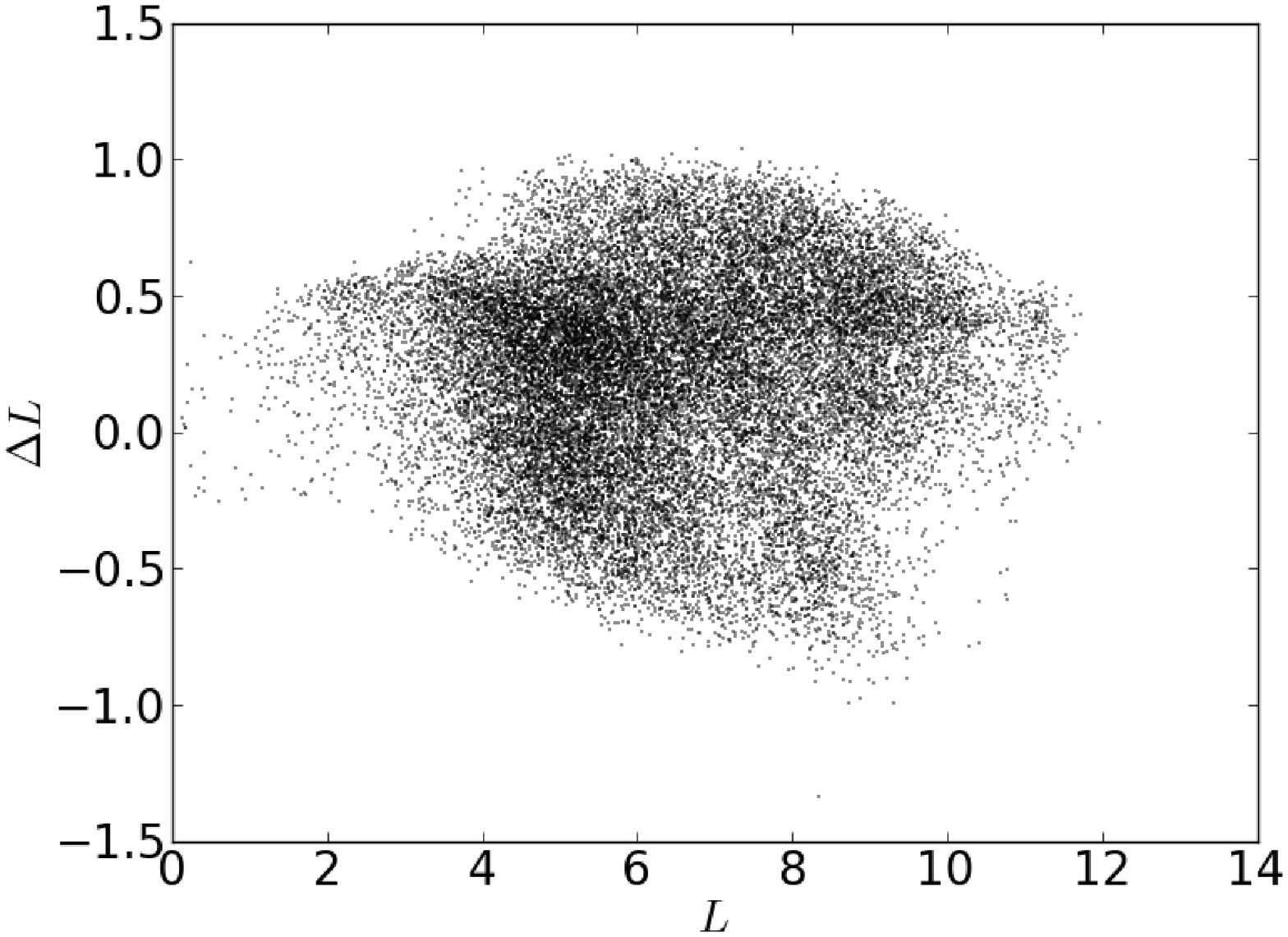}} \vspace{-6.0mm}\\
  \subfloat{\includegraphics[scale=0.28] {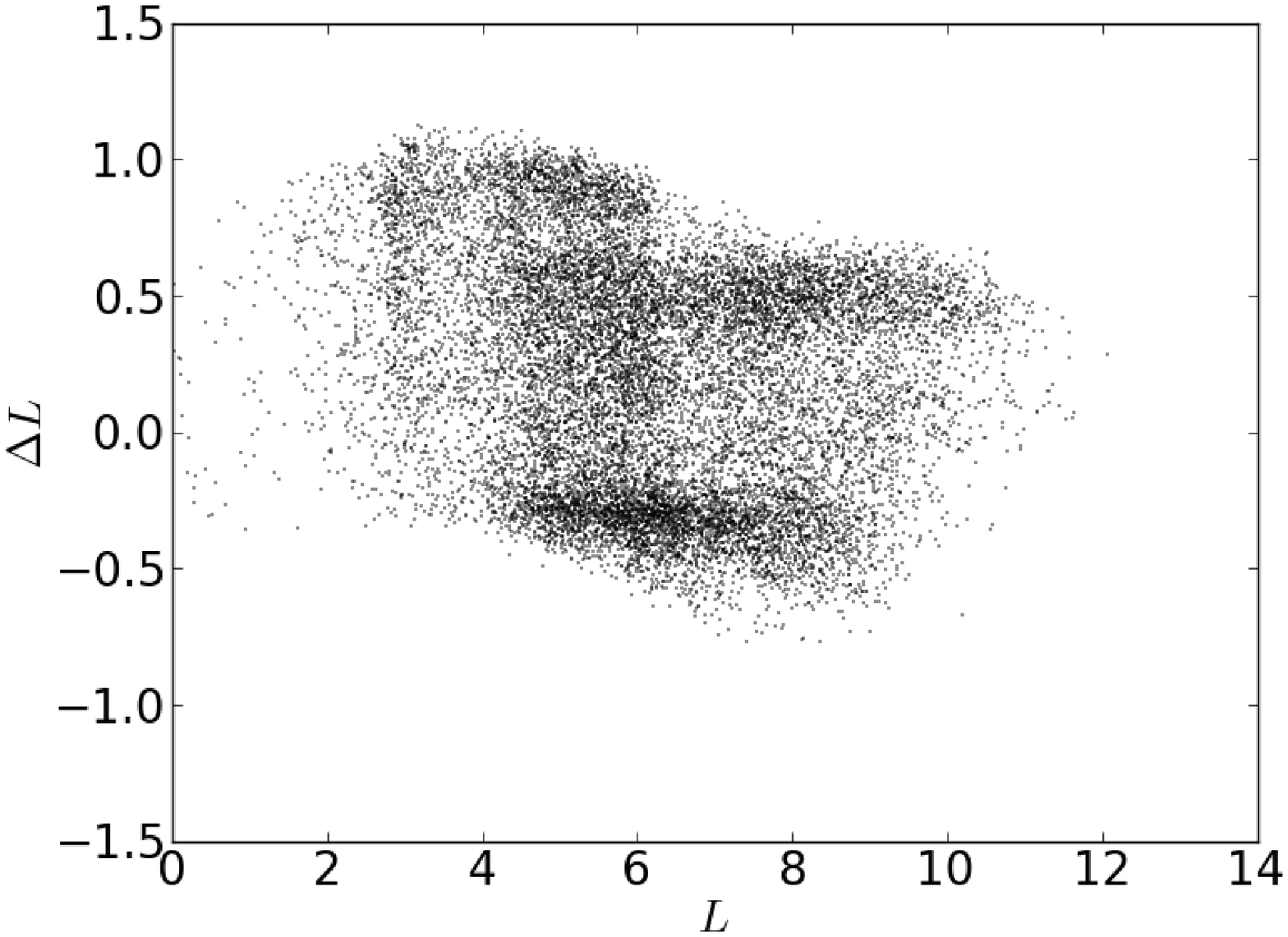}} \\

\caption[]{The same as Fig. \ref{LdLearly}, but at the late epoch of the weak bar. Each panel represents a stage of the spiral arm lifetime. From top to bottom: formation ($t=1.353$); fully formed single peak spiral arm ($t=1.393$); double peak spiral arm ($t=1.433$); breaking ($t=1.473$ Gyr). The trend is similar to that seen in the early epoch, but without the feature exhibiting large changes in angular momentum at lower initial angular momentum (seen in Fig. \ref{LdLearly}), which are induced by the strong bar.}
\label{LdLlate}
\end{center}
\end{figure} 

The same behaviour is observed at the early epoch as well. It is remarkable that although the spiral arm at the early epoch is systematically faster than the mean rotation velocity, we still observe these systematic migrations of star particles. This is probably because the pattern speed is not too different from the rotation velocity, and some star particles could be in a particular phase of their epicycle motion such that they are ripe for migration. Further studies of the orbits of these migrators are required, and will be studied in a forthcoming paper.

Fig. \ref{contlate} shows that the spiral arm is not always a strong single peak structure, owing to the winding and breaking of the spiral arm as it begins to disappear. The evolving structure of the spiral arm may affect the degree of radial migration that occurs over the stages of evolution that span from formation to destruction. Therefore, we select a new sample of stars over the whole spiral arm ($5 - 10$ kpc radius) within 2 kpc in the azimuthal direction from the expected arm position if the arm co-rotates with star particles around $t=1.034$ and $t=1.393$ Gyr i.e. the positions of the anchors shown in Figs. \ref{dentraceearly} and \ref{dentrace}. All selected particles are in the plane of the disc ($|z|<200$ pc) as before. This is done at the four stages of the spiral arm\textquotesingle s evolution: formation, single peak, double peak and finally destruction. In each case, the window of migration is 40 Myr, centred at each of these stages, which in total spans the lifetime of the spiral arm ($\sim 180$ and $130$ Myr for early and late epoch respectively).

The samples selected at the early epoch are plotted in the $L-\Delta L$ plane in Fig. \ref{LdLearly}. The largest migration occurs around the single peak stage when the arm is fully formed, and less migration occurs after this time when the double peak at $R \sim 6.5$ kpc develops. However, there appears to be a lot of negative migration at the stage of formation. This may be due to the tightly wound arm seen in the $t=0.978$ and $t=0.994$ Gyr panels in Fig. \ref{dentraceearly}. As seen from the anchor points, we sample the leading side of this arm and hence negative migrations are expected. There is also large migration present at low $L$, owing to the stronger bar at this early epoch. This procedure is repeated for the late epoch, and shown in Fig. \ref{LdLlate}. The same conclusions can be drawn for this weak bar case. As expected, the most migration occurs at the single peak stage when the arm is fully formed. At both epochs, significant migration occurs over a large range of radii. Furthermore, many panels show an obvious \textquotesingle two-pronged\textquotesingle \hspace{0.1mm} structure in the $L-$direction, one at positive $\Delta L$ and the other at negative $\Delta L$. This is a clear indication that radial migration occurs at a wide range of radii. The horizontal features are likely to be caused by a maximum $|\Delta L|$ in this short time period for migrating star particles along spiral arms.

\begin{figure}
\begin{center}
\subfloat{\includegraphics[scale=0.34]{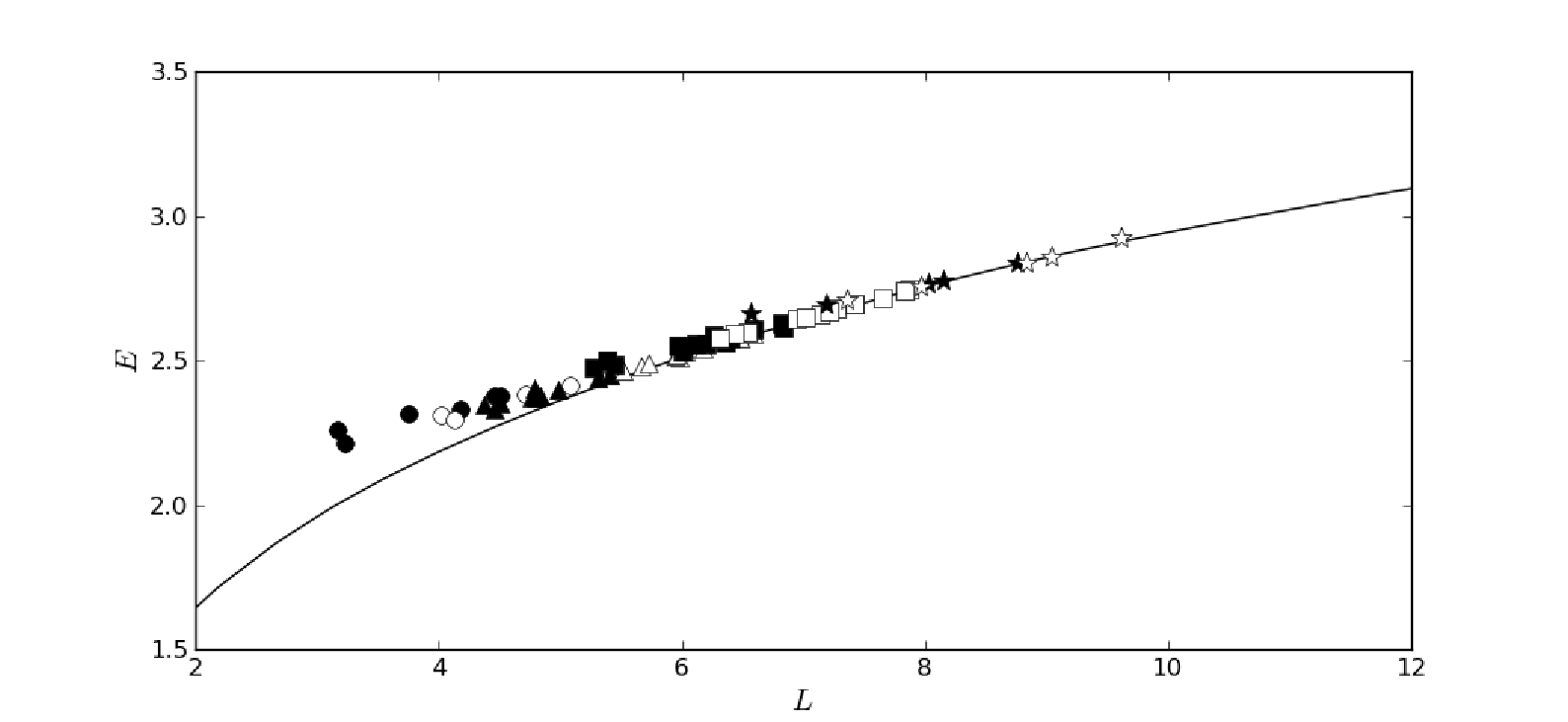}} \\ 
\subfloat{\includegraphics[scale=0.34]{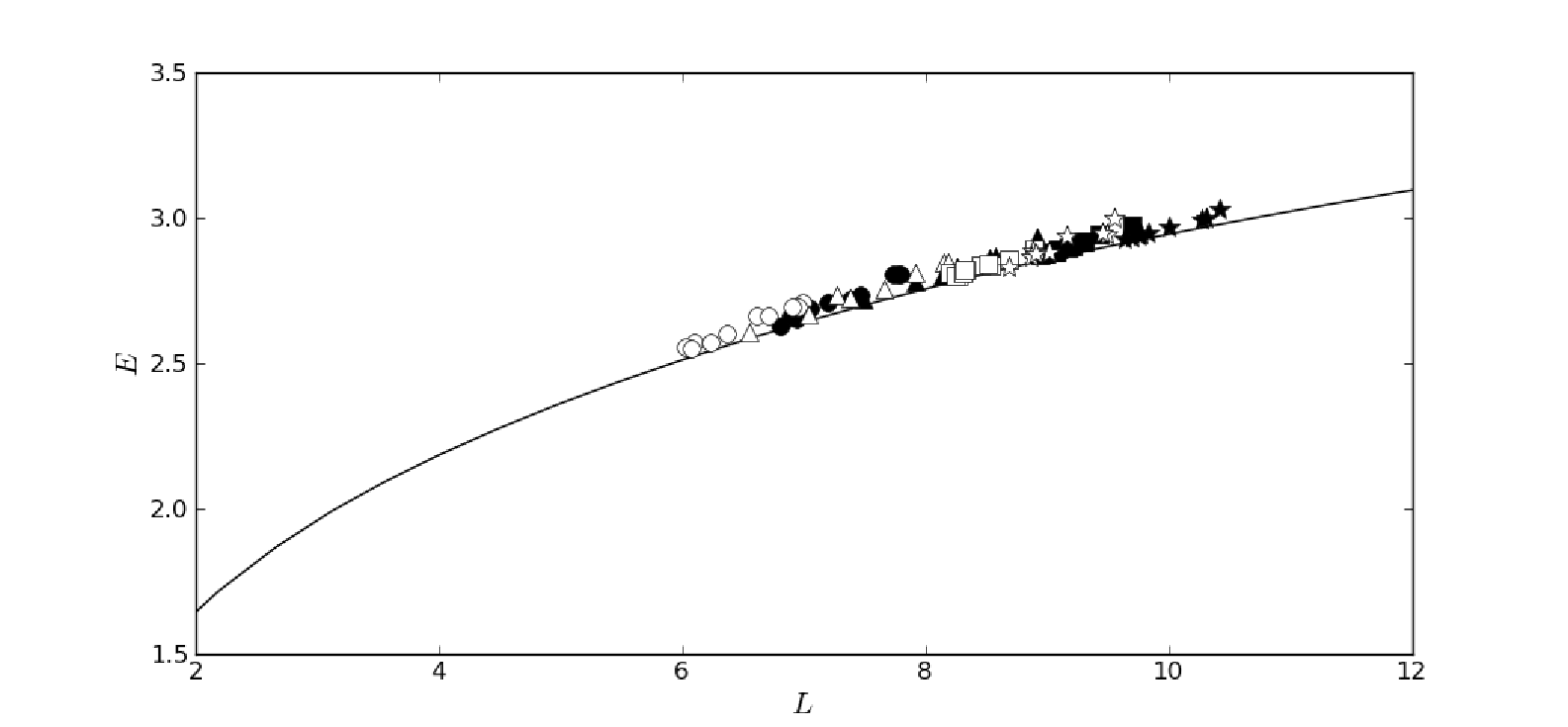}}\\
\caption[extvcdblpn]
{The energy, $E$, and angular momentum, $L$, distribution of the extreme migrators in Fig. \ref{5kpcplots} at 20 Myr before (filled symbols) and 20 Myr after (open symbols) the time step at which they were selected. Each symbol represents a specific radius of selection corresponding to Fig. \ref{5kpcplots}. The top (bottom) panel shows the results of the migrators that moved toward the outer (inner) radii. The solid black line indicates the circular orbit. Units are arbitrary.}
\label{extvcdblpn}
\end{center}
\end{figure}

\subsection{Angular momentum and energy evolution}

\begin{figure}
\begin{center}\hspace{-4.0mm}
\includegraphics[scale=0.38]{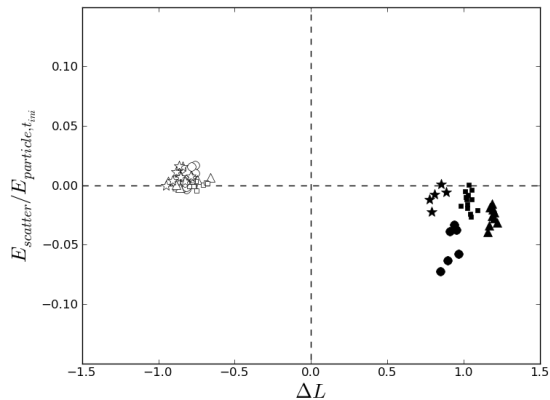}
\caption{The groups of migrators shown in Fig. \ref{extvcdblpn} are plotted here in the $\Delta L$ - $E_{scatter}/E_{particle,t_{ini}}$ plane, where $E_{scatter}$ is the change in the quantity defined by $E_{particle,t} - E_{circ}$ between the initial and final time. $E_{particle,t}$ is the total particle energy at a given time and $E_{circ}$ is the energy of a test particle of circular orbit for the given angular momentum (i.e. the minimum orbital energy allowed). $E_{scatter}/E_{particle,t_{ini}}$ tells us how much the star particle has gained or lost random energy as a fraction of the initial particle energy during the migration process. We can see the positive (filled symbols) and negative (open symbols) migrators lie in distinct groups, where the former are \emph{cooled} and the latter \emph{heated}, but only by a small amount.}
\label{migscat}
\end{center}
\end{figure}

\begin{figure*}
\centering
\includegraphics[scale=0.4]{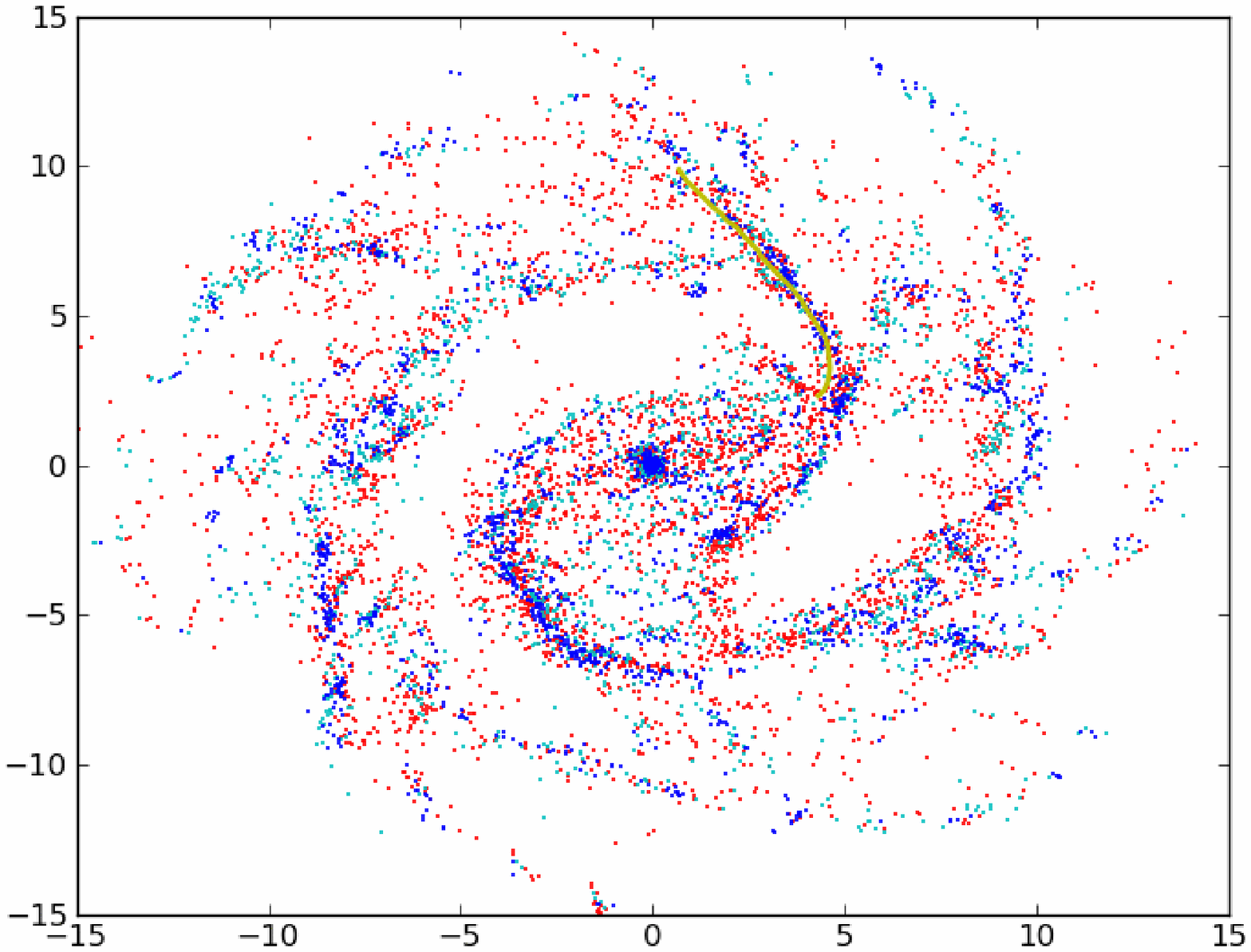}
\includegraphics[scale=0.4]{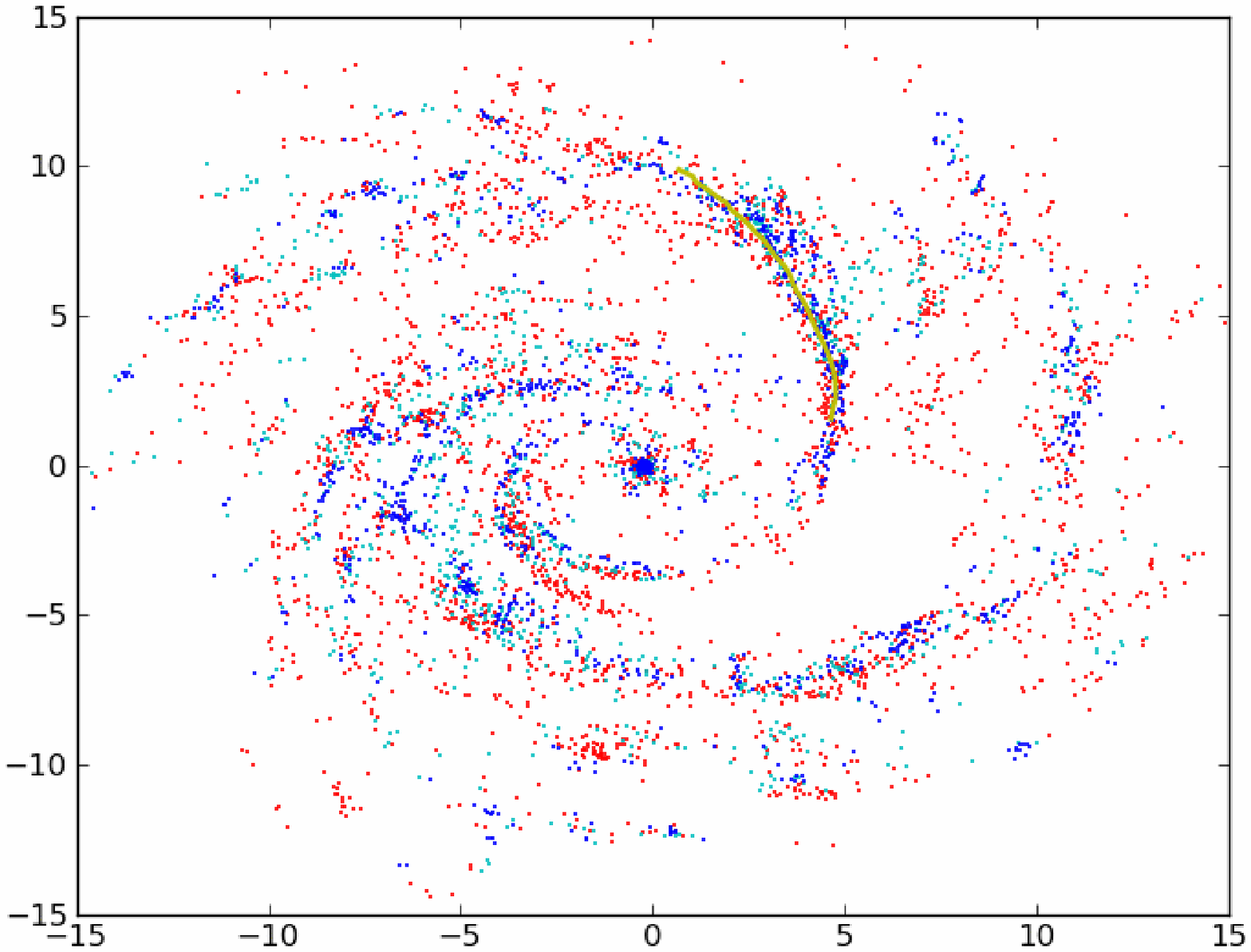}\\
\caption[]
{Snapshots of the disc at $t=1.034$ Gyr (\emph{left}) and $t=1.393$ Gyr (\emph{right}), showing only star particles of age: $t < 50$ Myr (blue); $50 < t < 100$ Myr (cyan); $100 < t < 200$ Myr (red). The yellow line indicates the stellar peak density of the spiral arm.}
\label{yngstars}
\end{figure*}

\begin{figure*}
\begin{center}
\subfloat{\includegraphics[scale=0.4]{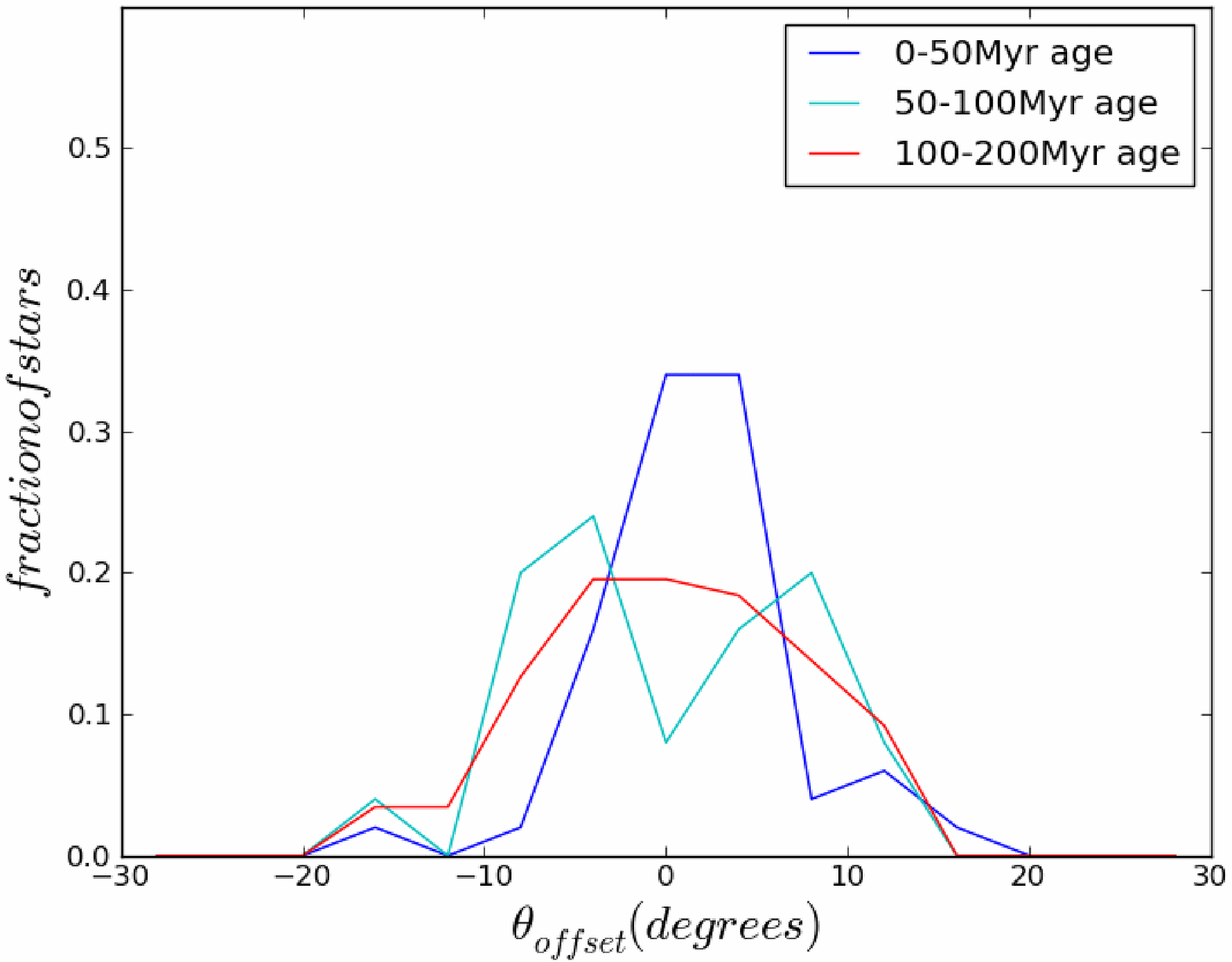}}
\subfloat{\includegraphics[scale=0.4]{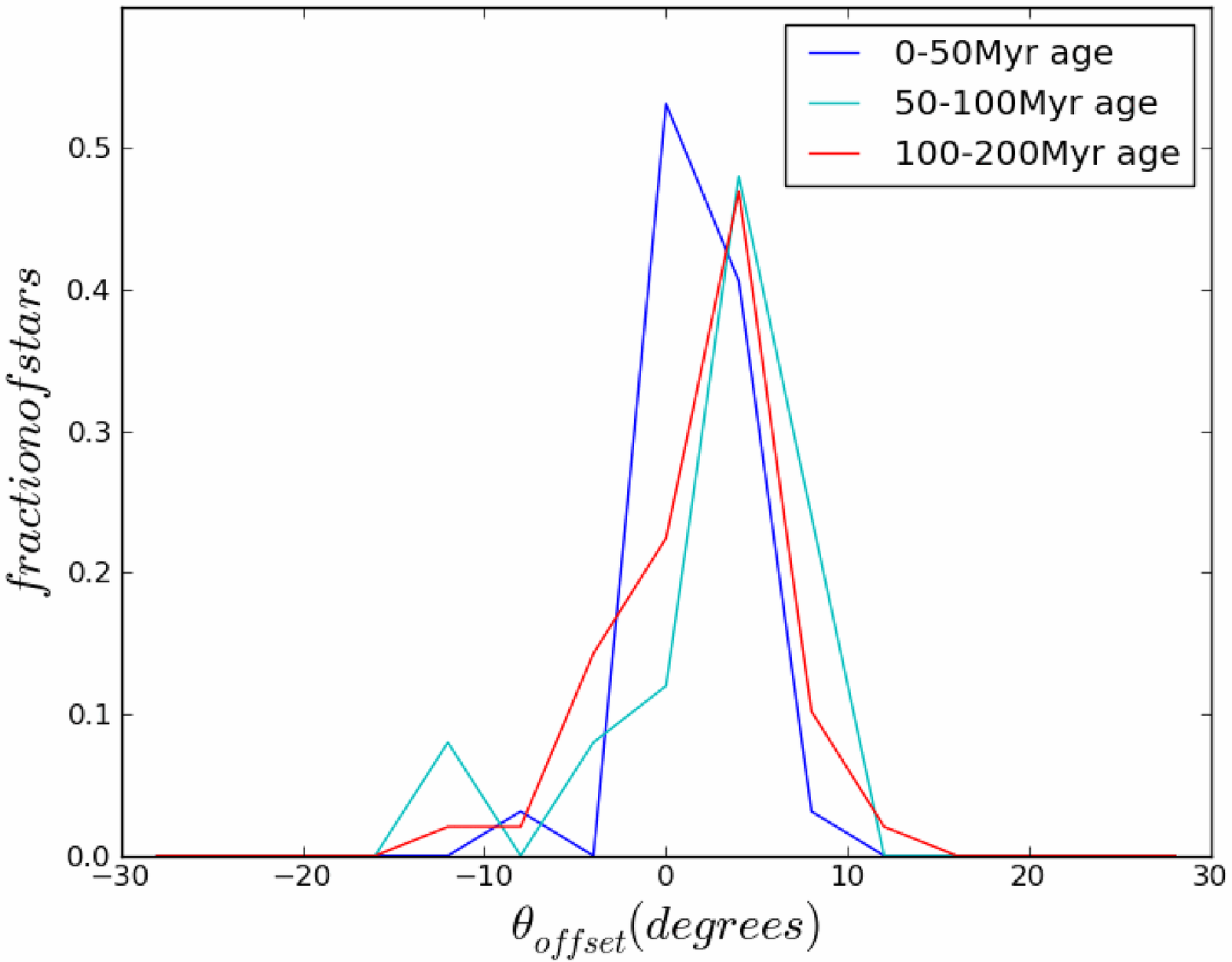}}\\
\subfloat{\includegraphics[scale=0.4]{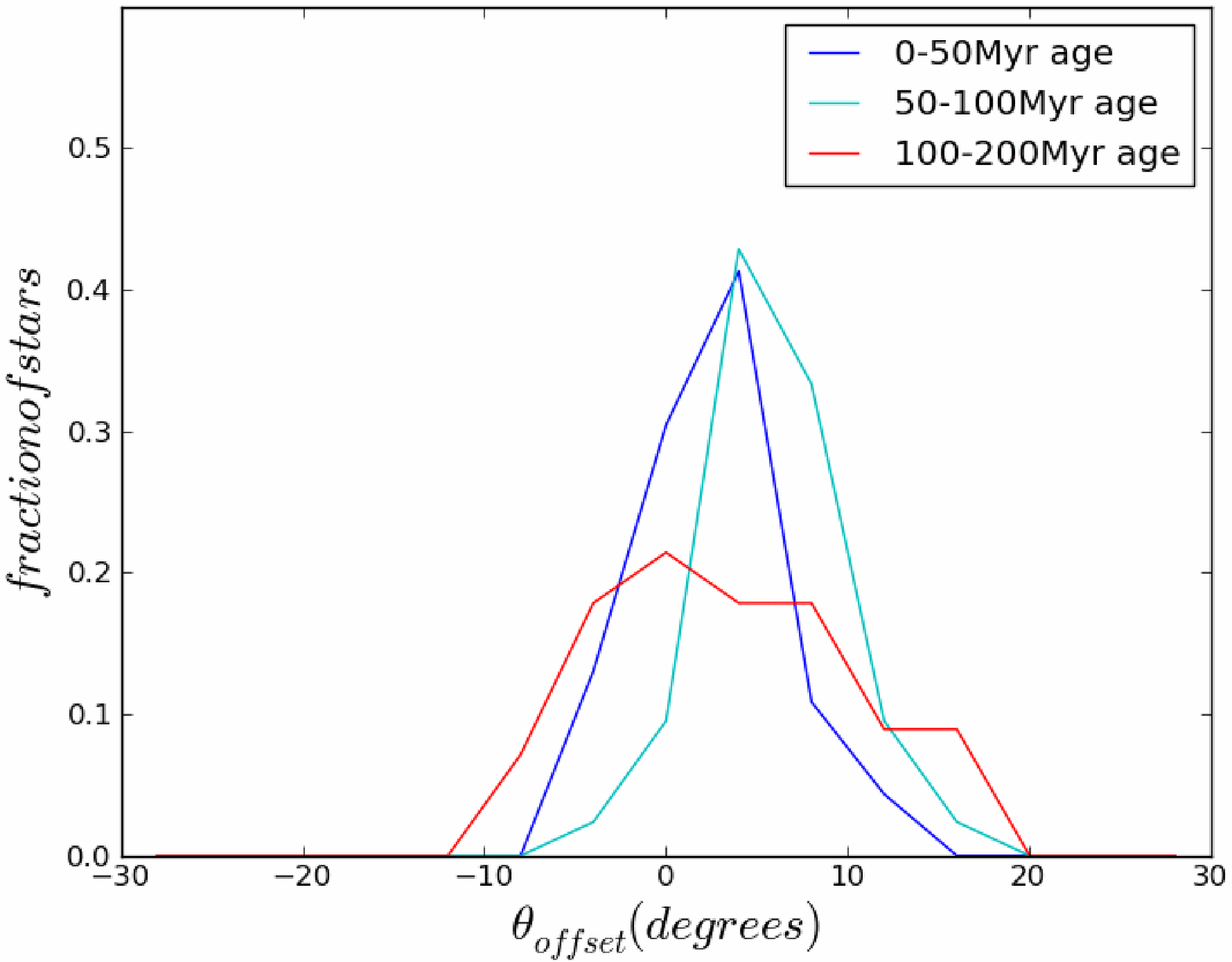}}
\subfloat{\includegraphics[scale=0.4]{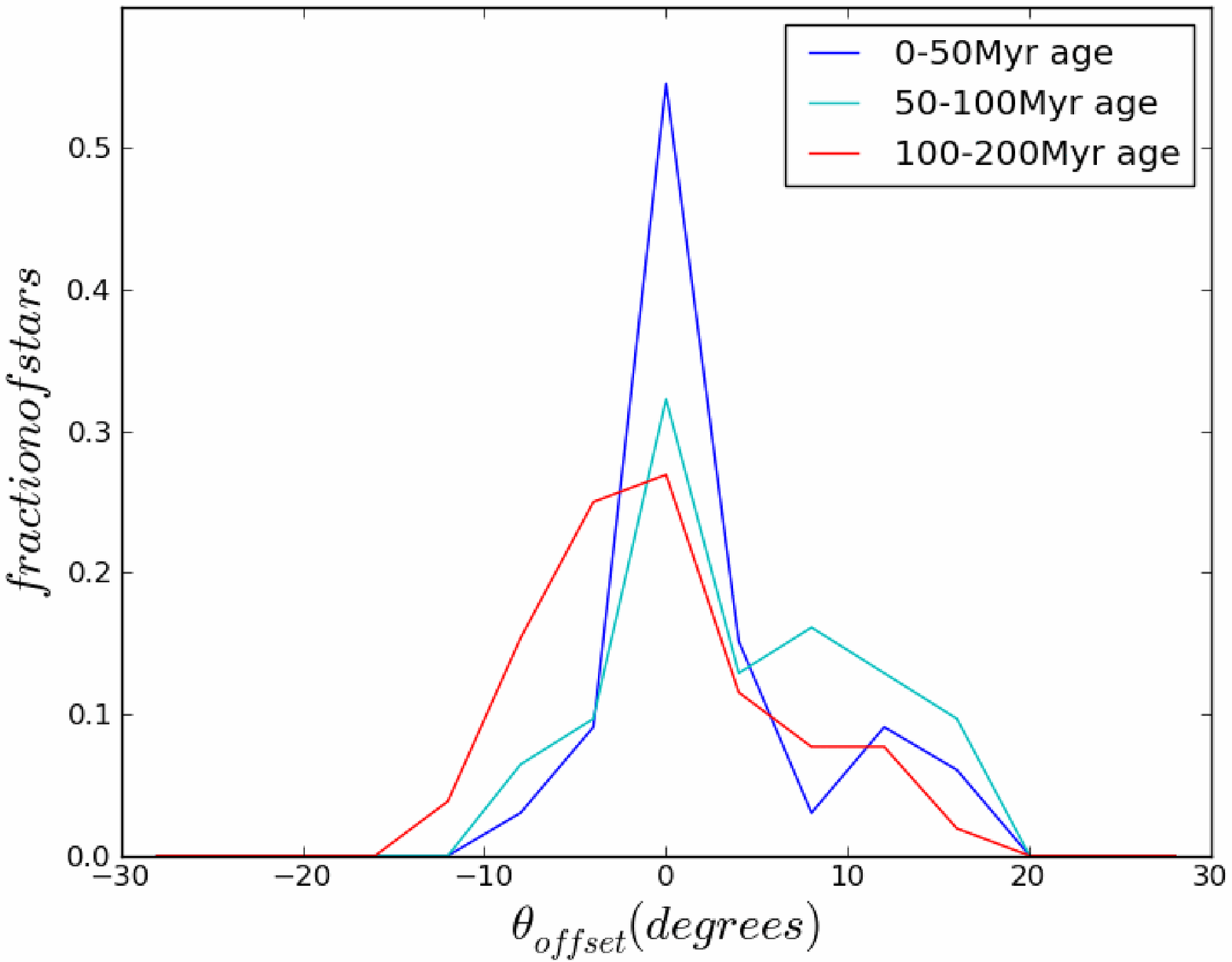}}\\
\caption[]
{Histogram of star particles at both early (top row) and late (bottom row) epochs. Particle samples are selected in the $6$ to $7$ kpc (left column) and $7.5$ to $8.5$ kpc (right column) radial range. Particles are binned according to their azimuthal offset from the peak density line shown in Fig. \ref{yngstars}. Negative offsets refer to the trailing side of the arm, and positive offsets refer to the leading side. No systematic offset of stellar ages is apparent.}
\label{offset1}
\end{center}
\end{figure*}

In this section, we focus on the angular momentum-energy evolution at the late epoch only (that of the earlier epoch is the same). The orbital energy of a star particle can be affected by the gain and loss of angular momentum associated with radial migration. As in \citet{GKC11} and following \citet{SB02}, we calculate the energy, $E$, and angular momentum, $L$, of the extreme migrators in Fig. \ref{5kpcplots} at 20 Myr before and after the time step at which they were selected (the top and bottom panels in Fig. \ref{5kpcplots}). We call these two time steps the \textquotesingle initial\textquotesingle \hspace{0.1mm} and \textquotesingle final\textquotesingle \hspace{0.1mm} time steps respectively. In Fig. \ref{extvcdblpn} we show the position of the extreme migrators at the initial (filled symbols) and final (open symbols) time steps for all migrator samples in Fig. \ref{deltallate}. The solid black line indicates the $L$ and $E$ expected for a pure circular orbit at each radius. This represents the minimum energy which a star particle can have at a given angular momentum. We see that the positive migrators (top panel) and negative migrators (bottom panel) move along the circular velocity curve in opposite directions to each other. Because they keep close to the circular velocity curve after migration, their orbits must retain their near circularity, and they gain (or lose) little random energy and are not scattered into higher energy orbits \citep{SB02}. In other words, their kinematically cool orbits are largely preserved. 

To quantify this, Fig. \ref{migscat} shows the amount of non-circular kinetic energy change (normalised to total initial energy) over the migration period plotted as a function of the amount of angular momentum change that they have undergone. Each migrator shows very little scatter during migration. It appears that each type of migrator is grouped separately, such that the positive migrators become slightly cooler, and negative migrators become slightly hotter\footnote{This is different from the global heating caused by scattering from spiral arms (see \citealt{Fu11}).} (see also \citealt{GKC11}; \citealt{RD11}), although this is less than a few percent of the initial energy.

\subsection{Stellar population distribution around spiral arms}

Because we have a gas component forming stars, we can now make a further test of whether the spiral arms are long-lived density waves as in spiral density wave theory. The long-lived, stationary wave theory should predict clear azimuthal offsets between young star particles of different ages, and molecular clouds (MCs) that are the seeds of star formation. The single constant pattern speed predicted by this theory would mean that inside the co-rotation radius, gas and stars would be moving faster than the spiral arm feature. Then gas flows into the spiral arm from behind the arm, and is compressed into MCs. This leads to star formation. Newly born stars will then flow through and begin to overtake the arm feature as they age, which naturally leads to a temporal gradient over the spiral arm. Outside of co-rotation, where material moves slower than the spiral arm feature, the opposite temporal gradient is expected. Therefore, if we group star particles around the spiral arm into age bins, and examine their azimuthal distribution there should be apparent azimuthal offsets among star particles of different ages, which would become clearer further from co-rotation. \citet{DB08} and \citet{DP10} performed a similar test by embedding a rigidly rotating spiral potential with a constant pattern speed.

We also analyse the azimuthal distribution of stellar ages found in our simulated galaxy. Note that although \citet{DP10} show similar analysis, they do not explicitly include radiative cooling or star formation, but assume an isothermal gas. They track the orbit of gas particles which have experienced the high density state, after which time the gas particles are tracked as very young stars ($2 - 100$ Myr), assuming the gas and stellar dynamics are similar in this short period. Therefore, our study is different and complementary. Fig. \ref{yngstars} shows the distribution of a young population, $ t_{age} < 50$ Myr (blue); an intermediate population, $ 50 < t_{age} < 100$ Myr (cyan); and a relatively older population, $ 100 < t_{age} < 200$ Myr (red).The snapshots shown is at $t = 1.034$ (left) and $t = 1.393$ (right) Gyr.

Inspection by eye indicates that there is no obvious offset between the tracers. To quantify this, we select two samples of star particles: one between $6$ and $7$ kpc radius and the other between $7.5$ and $8.5$ kpc radius, each within $\pm$ 2 kpc from the peak density of the spiral arm in the azimuthal direction. The angular offset distribution from the peak density for selected star particles of different ages are shown in Fig. \ref{offset1}, where the abscissa is azimuth offset angle and the ordinate is the number fraction of star particles. A negative angular offset is taken to mean a position behind the spiral arm, and a positive one means a position in front of the spiral arm. In both cases, no significant offset is seen between star particles of different ages.  Neither case finds any systematic spatial offset that would be present if the arm was a Lin-Shu type density wave. It is clear that the distribution broadens for older star particles, although the peak position remains about the same. Our results are qualitatively similar to the results of the flocculent and interacting galaxy cases in \citet{DP10}\footnote{Their barred galaxy case focuses on the stellar distribution around the bar not the spiral arm, and is therefore not relevant to our discussion in this section.}. As we expect, it is completely different from their fixed pattern speed case.

\section{Conclusions}

We have presented three dimensional N-body/SPH simulations of an isolated barred spiral galaxy, and performed a dynamical analysis of the spiral arms and particles around the spiral arms, tracing their evolution and the azimuthal distribution of star particles as a function of age. We come to the following conclusions:

\begin{enumerate}
\item{}
We find in our simulation that spiral arms are transient recurring features: we observe the continuous disappearance of spiral arms and the reappearance of new ones. This transient nature has always been found in numerical simulations.
\item{}
Our result shows that the pattern speed is decreasing with radius overall, and may be affected by the presence of a bar. The un-barred case shows convincing co-rotation with the rotational velocity. The weak bar case shows slight departure from rotational velocity at larger radii, and the stronger bar case shows a systematically faster pattern speed overall. Although we only studied three arms in this detail, this indicates that the bar may boost the pattern speed, and this deserves further study.
\item{}
It is demonstrated that despite the differences in pattern speed, each case exhibits the same systematic motion found in \citet{GKC11}, that leads to strong efficient migration.
\item{}
The spiral arms analysed are shown to develop a double peak substructure as it winds and evolves. The break occurs at the same radius at which the pattern speed kinks. The amount of radial migration is weaker at this stage and subsequent stages of the spiral arm evolution, although it still occurs until the spiral arm disappears. This is valid for both the weak and strong bar cases.
\item{} 
We quantify the amount of heating or cooling of each migrator in terms of random energy gained or lost over the process of migration. It is evident that each positive migrator loses some random energy (cools), while the negative migrators gain some random energy (heated). For each migrator, it is seen that the amount of heating/cooling is less than a percent of the total energy of a given particle. Hence this migration does not contribute significantly to disc heating. However the cause of this heating and cooling is not identified (see also \citealt{RD11}), and is worthy of further study.
\item{}
We find no offset between the distribution of young star particles ($< 200$ Myr) of different ages around the spiral arm at two different radii. This is consistent with recent observations of extra-galaxies (\citealt{FR11}; \citealt{FCK12}).
\end{enumerate}

This study is a follow-up study to our previous paper \citet{GKC11}, which focused on pure N-body simulations of a galaxy with no bar. As in that study, we have not addressed the mechanism of formation of the spiral arm features thoroughly nor their destruction, although we gain an insight into how the arm develops a double peak structure and then breaks. We note that the spiral arm features here are slightly longer-lived than our N-body galaxy, which could be because the bar is a powerful driver of spiral structure (e.g. \citealt{SS87}; \citealt{SLB10}), and may help to maintain the feature for longer (e.g. \citealt{DT94} \citealt{BT08}; \citealt{BAM09}, \citealt{QDBM10}). We also note that bars can be even stronger than the strong bar case presented here. It would be interesting to study the effects of spiral arm pattern speed on radial migration when the bar is much more prominent.

Again, we find that the spiral arms in this N-body/SPH barred galaxy are not consistent with the long-lived, rigidly rotating spiral arms of a classical spiral density wave theory. On the contrary, the spiral arm pattern speed decreases with radius and is similar to but slightly faster than the rotation velocity of the star particles. However, significant radial migration over a wide range of radii is repeatedly observed despite the differences in pattern speed. This suggests the existence of further criteria for radial migration, which will be studied in a forthcoming paper. Future studies should focus on testing this paradigm of the spiral arm with observations of our own Galaxy and of external galaxies (e.g. \citealt{MRM09}; \citealt{FR11}; \citealt{SW11}; \citealt{FCK12}), and comparing with simulations.

\section*{acknowledgements}
The authors acknowledge the support of the UK's Science \& Technology
Facilities Council (STFC Grant ST/H00260X/1). The calculations for
this paper were performed on Cray XT4 at Centre for
Computational Astrophysics, CfCA, of National Astronomical Observatory
of Japan and the DiRAC Facility jointly funded by STFC and the Large
Facilities Capital Fund of BIS. The authors acknowledge support of the
STFC funded Miracle Consortium (part of the DiRAC facility) in
providing access to the UCL Legion High Performance Computing
Facility.  The authors additionally acknowledge the support of UCL's
Research Computing team with the use of the Legion facility. 
This work was carried out, in whole or in part, through the Gaia
Research for European Astronomy Training (GREAT-ITN) network. The
research leading to these results has received funding from the
European Union Seventh Framework Programme ([FP7/2007-2013] under
grant agreement number 264895. The authors thank the referee for
a thorough inspection and important checks of the methodology.

\bibliographystyle{mn}
\bibliography{paper02-sphwb_jul12R2.bbl}

\end{document}